
%
%

\magnification=\magstep1
\openup 2\jot
\def\bo{ { \sqcup\llap{ $\sqcap$} } }
\overfullrule=0pt       


\font\list=cmcsc10

\def\real{I\negthinspace R}

\def\half{\textstyle{1\over2}}
\def\third{\textstyle{1\over3}}

\hbox{ }
\rightline {DAMTP/R-94/7}
\rightline {LA-UR-93-4473}
\rightline {March 1994}
\vskip 1truecm

\centerline{\bf THE INSTABILITY OF CHARGED BLACK STRINGS AND p-BRANES}
\vskip 1truecm

\centerline{Ruth Gregory}
\vskip 2mm

\centerline{ \it Department of Applied Mathematics and Theoretical Physics }
\centerline{ \it University of Cambridge, Silver St, Cambridge, CB3 9EW, U.K.}

\vskip 4mm

\centerline{ Raymond Laflamme }
\vskip 2mm

\centerline{ \it Theoretical Astrophysics, T-6, MSB288,
Los Alamos National Laboratory}
\centerline{ \it  Los Alamos,  NM 87545,USA}

\vskip 1.5cm
\centerline{\list abstract}
\vskip 3mm

{ \leftskip 10truemm \rightskip 10truemm

\openup -1\jot

We investigate the evolution of small perturbations around charged black strings
and branes which are solutions of low energy string theory.  We give the details
of the analysis for the uncharged case which was summarized in a previous paper.
We extend the analysis to the small charge case and give also an analysis for 
the generic case, following the behavior of unstable modes as the charge is 
modified.  We study specifically a magnetically charged black 6-brane, but 
show how the instability is generic, and that charge does not in general 
stabilise black strings and p-branes. 

\openup 1\jot

}

\vfill\eject
\footline={\hss\tenrm\folio\hss}

\noindent{\bf 1) Introduction}

Black holes have always been a source of fascination and debate, from the 
nature of the event horizon to the nature of the singularity and whether a 
full theory of quantum gravity can avert the apparent loss of unitarity in
black hole evaporation.
Classically,  four dimensional black holes are stable,  once
formed they settle down to a state described solely by their mass, charge and
angular momentum; therefore the singularities remain hidden from distant
observers.  
Quantum mechanically, black holes are quite different objects, 
analogous to a thermal system.  They have an entropy (proportional to
surface area) and a temperature, since like black bodies
they radiate thermally$^1$.  However, Hawking conjectured$^2$ that
a black hole formed from a pure quantum state would
radiate away leaving a mixed state
of radiation;   this would violate quantum mechanical unitarity.
It is certainly true that Hawking's semi-classical description will break 
down near the final stage of black hole evaporation as
planckian curvatures are approached, but
if quantum gravity preserves unitarity and information is to be
returned, it must do so well before this stage
 otherwise there is simply not enough energy left in a planck mass
black hole to emit all the information stored in a macroscopic black hole.
Therefore, in order to resolve this tension between quantum mechanics
and thermodynamics, it seems that
a semi-classical analysis of sub-Planck size black holes is needed.

Most recently, attention has focussed on low energy string gravity and its
implications for black holes. Some of these developments have been quite
interesting.  In Einstein gravity,
charged black holes (the Reissner-Nordstr\"om solutions), 
in addition to having an outer event horizon, have 
an inner Cauchy
horizon which is unstable to matter perturbations in the exterior spacetime$^3$.
However, there is no static charged black hole solution in Einstein gravity
with only one horizon and a spacelike singularity.
On the other hand, a key feature of low energy string gravity
is the presence of a dilaton, which greatly changes 
the causal structure of
charged black holes making them like Schwarzschild with one event horizon
and a spacelike singularity$^4$. This structure is generic, even if the
dilaton has a mass$^5$, as it must  to keep in line with the principle
of equivalence.

Of course, most of the analysis of stringy black holes has been performed 
in low dimensions, namely two or four, whereas string theory tells
us there should be ten dimensions, ideally therefore, one should be examining
black holes in ten dimensions. There has been work on black
holes in higher dimensions$^6$ for a range of horizon
topologies$^7$. In four dimensions, an event horizon must be topologically
spherical$^{8}$, but in higher dimensions this
is not necessarily the case, we
could have $S^2 \times $\real$^6$, or $S^3 \times $\real$^5$ topologies
for the horizon. In a previous letter we pointed out that a large
class of these black holes are unstable, namely the uncharged ones.
In this paper, we present the details of this original argument, as well
as providing an extension to cover charged black p-branes.

Why should black holes be stable, yet black strings, say, unstable? Before
answering this question in detail, it is worth examining a couple of naive 
arguments. Without loss of generality, consider a five dimensional black 
string in Einstein gravity, Sch$_4\times$\real. Then the equation
governing the metric perturbation $\delta g_{ab}=h_{ab}$, the Lichnerowicz
equation, is essentially a wave equation
$$
\Delta_L h_{ab} = (\delta^c_a \delta^d_b \bo + 2 R_{a\;b}^{\;c\;d}
)h_{cd} =0 .
\eqno (1.1)
$$
Because of the symmetries of the background Sch$_4\times$\real~metric, this
reduces to a four dimensional Lichnerowicz operator plus a 
$\partial_z^2$ piece. Performing a fourier decomposition of $h_{ab}$ in
the fifth dimension yields
$$
\Delta_L h_{ab} = (\Delta_4 - m^2)h_{ab}=0.
\eqno (1.2)
$$
Since the four dimensional Schwarzschild Lichnerowicz operator has no
unstable modes, adding a mass should only increase stability, hence it has 
been conjectured black strings are stable.

On the other hand, glibly speaking, horizons are like soap bubbles, they 
have a surface tension, $\kappa$ - the surface gravity, and soap bubbles do
not like being cylindrical! More formally,
a portion of horizon of length $L$ contains mass ${\cal M}=ML$
and has an entropy proportional to ${\cal M}^2/L$.
A five dimensional black hole on the other hand has
entropy proportional to ${\cal M}^{3/2}$. Thus for large lengths of horizon,
the mass contained within the horizon contributes a much lower entropy than
if it were in a hyperspherical black hole. This indicates that for large
wavelength perturbations in the fifth dimension, we might expect an
instability.

 As it turns out, the thermodynamic argument is correct. The flaw in the other
argument is to assume that the number of physical degrees of freedom of the
effective four dimensional tensor gauge field, $h_{ab}$, remains the 
same as in Einstein theory.  
Clearly this is not so, for the fifh dimension adds an effective mass
to $h_{ab}$.
A massive tensor field  in four dimensions has three
degrees of freedom as compared with only two for a massless one.
We will show that this extra degree
of freedom has a spherically symmetric mode which is responsible for the instability.

The stability of the five
dimensional black string was investigated analytically$^{9}$, with the
conclusion that there is no non-singular single unstable mode
on a Schwarzchild time $t=0$ surface, however, this argument did not 
{\it prove} stability.  As emphasized by Vishveshwara$^{10}$
in his
original Schwarzschild stability argument, the non-existence of a single
unstable mode does not preclude the existence of a composite unstable mode,
with the combination cancelling the singular behaviour of an inadmissable
single singular mode. An example of this sort of situation occurs in
the coloured black
hole instability, recently confirmed by Bizon and Wald$^{11}$.
That this was probably the case for five dimensional black strings was 
first pointed out by Whitt$^{12}$, who analyzed the stability of Schwarzschild
in four dimensional fourth order gravity - a different physical situation, 
but mathematically identical equations to
those studied in ref [9]. The key
difference in Whitt's analysis was the choice of an initial data
surface ending on the future horizon (see Figure 1.). By avoiding the neck of the
Schwarzschild wormhole, one avoids the fixed point of the isometries used to
generate the mode decomposition, which avoids in this case issues of
superposition. In any case, such a choice of initial data surface is mandated in
any real physical scenario, since a black hole or brane must form from collapse,
and  hence will not possess a Schwarzschild wormhole.

In a previous letter$^{13}$, we showed that using an appropriate initial data
surface, uncharged black $p$-branes were unstable, and conjectured that the
same would hold true for charged black $p$-branes. In this paper, we extend the
analysis of [13] to include charge,  and show that the instability does
indeed persist. It is worth stressing that this
instability is not of the Reissner-Nordstr\"om form - hidden behind the event
horizon, but it is a real {\it physical} instability of the exterior spacetime
which could potentially fragment the horizon. It is important to emphasize that
this can occur classically, for although under regular conditions horizons do not
bifurcate$^{14}$, if one has a naked singularity, then bifurcation is possible.
Since an instability calculation is by its nature linear, it cannot predict the
endpoint of an unstable evolution. However, the entropy argument does lend
support to the fragmentation scenario and violation of the Cosmic Censorship
Hypothesis.

The layout of the paper is as follows. In section two we summarise the black
p-branes we are investigating, namely those of Horowitz and Strominger$^7$, and 
derive the general perturbation equations. We discuss gauge constraints
and the issue of boundary conditions in detail, then we set up the mode
decomposition for our analysis. In section three we specialise to zero charge
and show how this greatly simplifies the problem, and present an instability.
We then use this `zeroth order' solution to generate an instability for small
charge in section four.  In section five we present the 
general case, for arbitrary charge, and present our conclusions in section
six.

\vskip 1cm

\noindent{\bf 2) The perturbation problem.}

We start this section on the derivation of the perturbation equations
by reviewing the particular black p-branes we will be investigating. We focus
on solutions of low energy string gravity with symmetry 
\real$^{10-D}\times$\real$_t\times$S$^{D-2}$ and a possible `magnetic'
charge. The action for such a theory is
$$
\int d^{10}x \sqrt{-g} e^{-2\phi}[ R + 4(\nabla \phi)^2 - {2\over(D-2)!}F^2]
\eqno (2.1)
$$
where $F$ is a $(D-2)$-form field strength. Varying this action yields the
equations of motion
$$
\eqalignno{
\bo \phi - 2 (\nabla\phi)^2 + F^2{(D-3)\over (D-2)!} &=0 & (2.2a) \cr
\nabla_{a_1}[e^{-2\phi} F^{a_1....a_{D-2}}] &=0 & (2.2b) \cr
R_{ab} + 2 \nabla_a \nabla_b \phi - {2\over (D-3)!}F_{aa_2...a_{D-2}}
F_b^{\;a_2....a_{D-2}} &=0 & (2.2c) \cr
}
$$
Searching for a solution with the required symmetry indicates a metric of the
form
$$
ds^2 = - e^A dt^2 + e^F dr^2 + e^B dx_idx^i + C^2 d\Omega^2_{D-2}
\eqno (2.3)
$$
where all metric functions are functions of the radial variable only,
and the index $i$ runs from $D+1$ to 10, and represents the coordinates
in the p-plane of symmetry of the p-brane. Looking for a `magnetically'
charged brane,
$$
F=Q\epsilon_{D-2}
\eqno(2.4)
$$
where $\epsilon_{D-2}$ is the area form on a unit (D$-$2)-sphere, it is
straightforward to show that a solution exists and takes the form$^7$
$$
\eqalign{
&e^A = {1-({r_+\over r})^{D-3} \over 1-({r_-\over r})^{D-3}}
\;\;\; , \;\;\; e^{-F} = \left ( 1-({r_+\over r})^{D-3} \right ) 
 \left ( 1-({r_-\over r})^{D-3} \right ) ,\;\;\;  B=0 \cr
&C(r) = r \;\;\; , \;\;\;
e^{-2\phi} =  1-({r_-\over r})^{D-3}\;\;\; , \;\;\;
Q^2 = {D-3\over 2} (r_+r_-)^{D-3} \cr
}
\eqno (2.5)
$$

If $r_-=0$, then $Q^2=\phi=0$, and the solution is the uncharged 
Sch$_D\times$\real$^{10-D}$, which we have argued to be unstable. However,
in the presence of charge there is also a non-trivial dilaton field, and it
is unclear how this will affect the physical instability. We first vary the
equations of motion to obtain the bare perturbation equations, then discuss
boundary
conditions on our perturbations in terms of a non-singular (or generalised
Kruskal) coordinate system on the event horizon, then return to the
perturbation equations, discussing gauge choices to simplify these,
finally discussing an additional simplification to the electromagnetic
perturbation.

In order to write the perturbation equations, we use the usual notation
$$
\delta g_{ab} = h_{ab}
\eqno (2.6)
$$
and $$
\bar{h}_{ab} = h_{ab} - \half h g_{ab}
\eqno (2.7)
$$
varying the equations of motion can be seen to give
$$
\eqalignno{
&\nabla_{a_1}[e^{-2\phi} (\delta F)^{a_1...a_{D-2}}] - 2 (\nabla_{a_1}
\delta\phi) e^{-2\phi} F^{a_1...a_{D-2}} - (D-3) e^{-2\phi}
F^{a_1b[...a_{D-2}} \nabla_{a_1} h_b^{a_2]} \cr
& \;\;\;\;\; - h^{a_1b}\nabla_{a_1} [ e^{-2\phi} F_b^{\;a_2...a_{D-2}}] 
-\nabla_{a_1}( {\bar h}^{a_1}_b) e^{-2\phi} F^{b....a_{D-2}}
=0  &(2.8a) \cr
& \bo \delta \phi - h^{ab}\nabla_a \nabla_b \phi +2 h^{ab} \nabla_a\phi
\nabla_b \phi - \nabla_c\phi  \nabla_b \bar{h}^{bc}
- 4 \nabla_a\delta \phi\nabla^a  \phi \cr
& \;\;\;\;\;  -{h^{cd}\over (D-4)!}
F_{ca_2...a_{D-2}} F_d^{\;a_2...a_{D-2}} + 2 {(D-3) \over (D-2)!}
\delta F_{a_1...a_{D-2}} F^{a_1...a_{D-2}}=0 & (2.8b) \cr
&\bo h_{ab} + 2 R_{cadb}h^{cd} - 2 R_{e(a} h_{b)}^e - 2 \nabla_{(a} \nabla_{|e|}
\bar{h}^e_{b)} - 4 \nabla_a \nabla_b \delta\phi - 2 \nabla_c\phi \nabla^c h_{ab}
+ 4 \nabla_c \phi \nabla_{(b} h_{a)}^c \cr
& + {4\over (D-3)!} \left [ 
2 F_{(a}^{\;\;a_2...a_{D-2}} \delta F_{b)a_2...a_{D-2}} - (D-3) h^{cd}
F_{aca_3...a_{D-2}} F_{bd}^{\;\;a_3...a_{D-2}} \right ] =0 & (2.8c) \cr
}
$$
These are the bare perturbation equations, we must now examine boundary
conditions, and whether there is any
choice of gauge which simplifies (2.8).

\noindent{\it Boundary conditions.}

Now we turn to the question of boundary conditions, which
are a key to this problem. Obviously, we want to place initial
data on a Cauchy surface for the exterior spacetime, but such a surface
necessarily touches the horizon, which is singular in Schwarzschild
coordinates. There are therefore two issues here: One is 
how to define `small' for the perturbation at the horizon,
and secondly, which initial
data surface to impose these constraints upon.

The first issue is straightforwardly dealt with. 
Although the horizon is
singular in Schwarzschild coordinates, it is not a physical singularity,
merely a coordinate singularity. In four dimensions, non-singular
coordinates have been known for some time - Kruskal coordinates. In order
to generalise these to higher dimensions, the starting point is to identify
the generalised tortoise coordinate
$$
\eqalignno{
r^*_D &= \int e^{(F-A)/2} dr = \int {r^{D-3}\over (r^{D-3} - r^{D-3}_+)}dr
=\sum_{n=1}^{D-3} {1\over D-3} \int {dr \over (1-e^{2\pi in\over D-3 } r_+/r)} \cr
&= r + \sum_{n=1}^{D-3} {e^{2\pi in\over D-3} \over (D-3)} r_+ \log
  (r-e^{2\pi in\over D-3} r_+)&(2.9) \cr
}
$$
So, for example,
$$
\eqalign{
r_4^* &= r + r_+ \log (r-r_+) \cr
r_5^* &= r + \half r_+ \log {(r-r_+)\over (r+r_+)}  \cr
r_6^* &= r + \third r_+ \log (r-r_+) - {r_+\over 6}\log(r^2+rr_+ +r_+^2)
+{r_+\over \sqrt{3}} \tan^{-1} \left ( {\sqrt{3}r_+\over 2r+ r_+}
\right ) \cr
}
\eqno(2.10)
$$
etc.
In particular, as $r\to r_+$
$$
r_D^* \sim {r_+\over D-3} \log (r-r_+)
\eqno(2.11)
$$

Having the tortoise coordinate, we can now define the Kruskal null coordinates
$$
\eqalign{
P_+ &= \exp \{ (D-3) (t+r^*)/2r_+ \} \cr
P_- &= -\exp \{ -(D-3)(t-r^*)/2r_+ \} \cr
}
\eqno (2.12)
$$
In terms of which the radial-time part of the metric looks like
$$
ds_2^2 = {4r_+^2 (D-3)^2 e^A \over P_+ P_-} dP_+ dP_-
\eqno (2.13)
$$
which is finite as $r\to r_+$.

Finally, we set
$$
\eqalign{
R &= P_+ - P_- \cr
T &= P_+ + P_- \cr
}
\eqno (2.14)
$$
in terms of which
$$
ds^2_2 = - {r_+^2 (D-3)^2 e^A \over P_+ P_-} (dR^2 - dT^2) 
= f^2(r)(dR^2 - dT^2)
\eqno (2.15)
$$
in order to readily identify our initial data surface.

This leaves us with the problem of an initial data surface. The domain
of dependence must obviously include ${\cal I}^+$, thus a surface touching
the future horizon, or the neck of the Schwarzschild wormhole is acceptable,
but a surface touching the past horizon is not, unless it passes through and
extends to the opposite horizon on the Penrose diagram. Following Whitt,
we impose regularity in the Kruskal system as $r\to r_+$ at constant
non-zero $T$, i.e.,, $R\to T$ from above as depicted on Figure 1. This
avoids the issue of mode superposition discussed earlier, and secondly, we
believe it to be a better physically motivated choice of surface. This is
because in practice a black hole (or brane) would form in a collapse
situation, and hence would not have a Schwarzschild wormhole; analyzing
stability would necessarily require a surface ending on a future event
horizon.

In terms of the Schwarzschild components
$$
\eqalignno{
h^{TT} &= - {(D-3)^4 \over f^2(r) (T^2-R^2)}
\left [ R^2 h^{\hat{t}\hat{t}} + 2RT h^{\hat{t}\hat{r}} 
+ T^2 h^{\hat{r}\hat{r}} \right ] & (2.16a) \cr
h^{RR} &= - {(D-3)^4 \over f^2(r) (T^2-R^2)}
\left [ T^2 h^{\hat{t}\hat{t}} + 2RT h^{\hat{t}\hat{r}} 
+ R^2 h^{\hat{r}\hat{r}} \right ] & (2.16b) \cr
h^{TR} &= - {(D-3)^4 \over f^2(r) (T^2-R^2)}
\left [ RT (h^{\hat{t}\hat{t}} + h^{\hat{r}\hat{r}}) 
+ (T^2+R^2) h^{\hat{t}\hat{r}} \right ] & (2.16c) \cr
h^{T*} &= {(D-3)^2 \over f(r) \sqrt{(R^2-T^2)}}
\left [ R h^{\hat{t}*} + T h^{\hat{r}*} \right ] & (2.16d) \cr
h^{R*} &= {(D-3)^2 \over f(r) \sqrt{(R^2-T^2)}}
\left [ T h^{\hat{t}*} + R h^{\hat{r}*} \right ] & (2.16e) \cr
}
$$
all of which should be finite as $r\to r_+$ on our initial data surface.
Since $R=T+O(r-r_+)$ as $R \to T$ from above, this implies that the various 
combinations of the normalised $h^{{\hat a}{\hat b}}$ within the square brackets
above must be $O(r-r_+)$ as we approach the event horizon. Now we turn to
simplifying the perturbation equations by a judicious choice of gauge.

\noindent{\it Gauge considerations.}

In gravity, physics must be invariant under a re-labelling of
coordinates. Such general coordinate
transformations (gct's), are generated by vector fields $\xi^a$,
the effect of an infinitesimal gct being to push the coordinates an 
infinitesimal amount
along the integral curves of $\xi^a$, such that $x^a\rightarrow x^a + \xi^a$.
Under such a gauge transformation, physical quantities  transform as
$$
\cal {P} \to \cal{L}_{\xi} \cal{P}
\eqno (2.17)
$$
where $\cal{L}_\xi$ denotes the Lie derivative with respect to $\xi$.
Therefore a pure gauge perturbation is of the form
$$
\eqalign{
h_{_\xi ab} &= 2\xi_{(a;b)} \cr
\delta \phi &= \xi^a \nabla_a \phi \cr
\delta F_{a_1...a_{D-2}} &=  F_{a_1...a_{D-2};d} \xi^d + 
(D-2) F_{d[a_2...a_{D-2}} \xi^d_{;a_1]} \cr
}
\eqno (2.18)
$$
It is straightforward to verify that such a pure gauge perturbation satisfies
our generalised Lichnerowicz perturbation equations. 

Using such a pure gauge perturbation, we see that
$$
\nabla_a \bar{h}_\xi^{ab} = R^b_a \xi^a + \bo \xi^b
\eqno (2.19)
$$
However, the right hand side of this equation is the curved space wave
operator, which given a source and initial conditions can be integrated
to give a solution for $\xi^a$ throughout the manifold$^{15}$. Thus, if the 
divergence of ${\bar h}_{ab}$ is not initially zero, we can change  gauge
in order to make it so. This simplifies (2.8) to the following:
$$
\eqalignno{
&\nabla_{a_1}[e^{-2\phi} (\delta F)^{a_1...a_{D-2}}] - 2 (\nabla_{a_1}
\delta\phi) e^{-2\phi} F^{a_1...a_{D-2}} - (D-3) e^{-2\phi}
F^{a_1b[...a_{D-2}} \nabla_{a_1} h_b^{a_2]} \cr
& \;\;\;\;\; - h^{a_1b}\nabla_{a_1} [ e^{-2\phi}
 F_b^{\;a_2...a_{D-2}}] =0  &(2.20a) \cr
& \bo \delta \phi - h^{ab}\nabla_a \nabla_b \phi +2 h^{ab} \nabla_a\phi
\nabla_b \phi
- 4 \nabla_a\delta \phi\nabla^a  \phi \cr
& \;\;\;\;\;  -{h^{cd}\over (D-4)!}
F_{ca_2...a_{D-2}} F_d^{\;a_2...a_{D-2}} + 2 {(D-3) \over (D-2)!}
\delta F_{a_1...a_{D-2}} F^{a_1...a_{D-2}}=0 & (2.20b) \cr
&\bo h_{ab} + 2 R_{cadb}h^{cd} - 2 R_{e(a} h_{b)}^e 
 - 4 \nabla_a\nabla_b \delta\phi - 2 \nabla_c\phi \nabla^c h_{ab}
+ 4 \nabla_c \phi \nabla_{(b} h_{a)}^c \cr
& + {4\over (D-3)!} \left [ 
2 F_{(a}^{\;\;a_2...a_{D-2}} \delta F_{b)a_2...a_{D-2}} - (D-3) h^{cd}
F_{aca_3...a_{D-2}} F_{bd}^{\;\;a_3...a_{D-2}} \right ] =0 & (2.20c) \cr
}
$$
The residual gauge freedom are those gct's which 
satisfy
$$
R^b_a \xi^a + \bo \xi^b = 0
\eqno (2.21)
$$
Notice that in a vacuum spacetime, i.e.~no dilaton, no `electromagnetism',
the trace of the metric perturbation equation becomes
$$
\bo h=0
\eqno (2.22)
$$
and we can choose an harmonic $\xi^a$ to set $h=0$. Therefore {\it in vacuo}
we may make the additional gauge choice of trace-free. However, more
generally it should be noted that tranversality is all that can be assumed.

Since a pure gauge perturbation automatically satisfies the perturbation 
equations, deciding whether a putative instability is physical or not
reduces to investigating whether it can be expressed in this form. 
Alternatively, one can identify physical perturations by using gauge
invariant variables.  It is straightforward to adopt Bardeen's results$^{16}$
to the case at hand.  The idea is to choose a particular linear combination
of variables and their derivatives such that the result becomes independent
of $\xi^a$.  The details and a particular set of gauge independent variables
are given in Appendix A.

  Finally, before beginning the analysis of the equations in detail we will
make a further simplification by showing that the `electromagnetic'
perturbation, $\delta F_{ab}$, can be taken to be zero. In order to perform a
stability analysis we will Fourier decompose all the perturbations in terms of
the symmetries of the background spacetime. Since Kruskal coordinates are not
well suited to such a decomposition, we perform our analysis in
Schwarzschild coordinates,  transforming
to Kruskal at the horizon to check boundary conditions according to (2.16).
 The Fourier modes in the time and $p$-brane directions are
easy to identify, for an instability they  are of the form $e^{\Omega t
+i\mu_ix^i}$. The spherical harmonic modes will depend on the number of
dimensions, $D$, that the black hole sits in, as well as on the tensorial nature
(scalar, vector, and so on) of the perturbation being analysed.
However, since in general higher angular momentum modes are more stable,
and since our advertised instability is an s-mode, zero angular
momentum modes are all we shall be considering. This clearly means we can
make no stability claims if we fail to find an instability, however, since
our primary concern is to chart regions of instability, an s-wave
analysis will suffice.

Therefore, the spherically symmetric perturbations take the form:
$$
\eqalignno{
\delta\phi &= e^{\Omega t+i\mu_ix^i} f(r) & (2.23a) \cr
\delta F &= e^{\Omega t+i\mu_ix^i} \cases{
q(r) \epsilon_{D-2} & angular components \cr
0 & mixed angular / t-r-i components \cr
\delta F_T & t-r-i components \cr }
& (2.23b) \cr
h^{ab} &= e^{\Omega t+i\mu_ix^i} \left [
\matrix{ H^{ij}(r)& H^{it} &H^{ir}& 0 &0 &... \cr
         H^{tj} & H^{tt} & H^{tr} & 0 &0&...\cr
         H^{rj} & H^{tr} & H^{rr} & 0 &0&...\cr
		       0 & 0 & 0 & K(r)&0&...\cr
	        0 & 0 & 0 & 0 &K/\sin^2\theta &...\cr
          ...&...&...&...&...&...\cr} \right ] & (2.23c) \cr
}
$$
Where $\delta F_T$ can be non-zero only if $D-2 \leq 5$. 

We may now see that the $F$-perturbation equation, (2.20b), becomes simplified,
for noting that the background field $F = Q\epsilon_{D-2}$ has only angular
components, and that $h^\alpha_\beta = K(r) \delta^\alpha_\beta/r^2$ for
angular components, 
$$
F^{a_1b[...a_{D-2}}\nabla_{a_1} h_b^{a_2]} =
F^{a_1b[...a_{D-2}}h^{a_2]}_{b,a_1} + F^{a_1b[...a_{D-2}} \Gamma^{a_2]}_{ca_1}
h^c_b = 0
$$
and
$$
\nabla_{r/t}[e^{-2\phi} F_{r/t}^{a_2...a_{D-2}}] = 0 \Rightarrow 
h^{a_1b}\nabla_{a_1}[e^{-2\phi} F_b^{a_2...a_{D-2}}] \propto g^{ab} 
\nabla_{a_1}[e^{-2\phi} F_b^{a_2...a_{D-2}}] =0
$$
Hence the $F$-perturbation equation reduces to
$$
\nabla_{a_1}[e^{-2\phi} (\delta F)_b^{a_1...a_{D-2}}] =0
\eqno(2.24)
$$
We can make the further observation that since $F$ and $\delta F$ are derived
from a potential,  $d(\delta F) =0$, hence
$$
\delta F_{\theta\phi.....,t} = \Omega q(r) \epsilon_{\theta\phi.....} = 0
\eqno (2.25)
$$
provided $\Omega\neq 0$. Thus only $\delta F_T$ can be non-zero, and this is
independent of the angular variables.

We may now argue that  
$\delta F_T$ also vanishes. Without loss of generality, we will show this for
$D=4$, the generalisation to higher dimensions being an iterative process.

Consider $d(\delta F) =0$. 
$$
F_{ab,c} + F_{bc,a} + F_{ca,b} = 0
\eqno (2.26)
$$
For $a=t$, $b=r$, $c=i$, we have
$$
\delta F_{_Tit}' + \Omega \delta F_{_Tri} + i\mu_i\delta F_{_Ttr} = 0
\eqno (2.27)
$$
Setting $f_t = \Sigma \mu_i\delta F_{_Tit}$ and $f_r = \Sigma \mu_i\delta
F_{_Tir}$ and summing (2.27) appropriately gives
$$
f_t' - \Omega f_r + i\mu^2\delta F_{_Ttr} = 0
\eqno (2.28)
$$
where $\mu^2 = \Sigma \mu_i^2$. Now, we return to the equation of motion (2.24)
which implies
$$
\Omega \delta F_T^{ta} + \sum i\mu_i \delta F_T^{ia} + \delta F_T^{ra\prime} +
{2\over r} \delta F_T^{ra} = 0
\eqno (2.29)
$$
For a=r, we have
$$
-{(r-r_-)\over (r-r_+)} \Omega\delta F_{_Ttr}  +if_r = 0
\eqno (2.30)
$$
Substituting from (2.28) implies
$$
f_t' = \left [ {\mu^2(r-r_+)\over \Omega (r-r_-)} + \Omega \right ] f_r
\eqno (2.31)
$$
Whereas taking (2.29) for $a=i$, multiplying by $\mu_i$ and summing gives
$$
-\Omega g^{tt} f_t - (g^{rr}f_r)' - {2\over r} g^{rr} f_r = 0
\eqno (2.32)
$$
Substituting for $f_r$ from (2.31), and rearranging gives finally
$$
\eqalign{
{(r-r_-)(r-r_+)\over r^2}f_t'' 
&+\left ( {(2r-r_+-r_-)\over r^2} 
     - {\mu^2(r_+-r_-)(r-r_+)\over r[\mu^2(r-r_+)+\Omega^2(r-r_-)]}
 \right )f_t'  \cr
&- \left(\mu^2 + \Omega^2{(r-r_-)\over (r-r_+)} \right )f_t = 0 \cr}
\eqno (2.33)
$$
By inspection of this equation we can see that there are no regular solutions
and thus we must take $f_t =0$. Hence $f_r=  \delta F_{_Ttr}=0$.

Now, returning to (2.27) implies
$$
\delta F_{_Tit}' + \Omega \delta F_{_Tri} = 0
\eqno (2.34)
$$
and (2.26) for $a=i$, $b=t$, $c=j$ implies
$$
i\mu_j \delta F_{_Tit} + i\mu_i \delta F_{_Ttj} + \Omega \delta F_{_Tji} =0
\eqno (2.35)
$$
Multiplying by $\mu_i$ and summing gives
$$
i\mu^2 \delta F_{_Ttj} - \Omega f_j = 0
\eqno (2.36)
$$
where $f_j = \Sigma\mu_i \delta F_{_Tij}$. Taking the $j$-component of the
equation of motion
$$
\Omega \delta F_{_Ttj}g^{tt} + if_j + (\delta F_{_Trj}g^{rr})' + 
{2\over r}\delta F_{_Trj}g^{rr} = 0,
\eqno (2.37)
$$
substituting in for $f_j$ , $\delta F_{_Trj}$, and rearranging gives
$$
{(r-r_-)(r-r_+)\over r^2}\delta F_{_Ttj}'' 
+ {(2r-r_+-r_-)\over r^3} \delta F_{_Ttj}' 
- \left(\mu^2 + \Omega^2{(r-r_-)\over (r-r_+)} \right )\delta F_{_Ttj} = 0 
\eqno (2.38)
$$
As before, this has no regular solutions, hence $\delta F_{_Ttj}=0$. (2.34) and
(2.35) then imply that all other components of $\delta F_T$ vanish, hence the
perturbation $\delta F_{ab}$ may be taken to be zero in all future analysis of
the spherically symmetric perturbation equations.

Finally, to summarize, the equations of motion for the spherically symmetric
perturbations we are interested in analysing reduce to the following 
 $$
\eqalignno{
& \bo \delta \phi - h^{ab}\nabla_a \nabla_b \phi +2 h^{ab} \nabla_a\phi
\nabla_b \phi
- 4 \nabla_a\delta \phi\nabla^a  \phi \cr
& \;\;\;\;\;  -{h^{cd}\over (D-4)!}
F_{ca_2...a_{D-2}} F_d^{\;a_2...a_{D-2}} =0 & (2.39a) \cr
&\bo h_{ab} + 2 R_{cadb}h^{cd} - 2 R_{e(a} h_{b)}^e 
 - 4 \nabla_a\nabla_b \delta\phi - 2 \nabla_c\phi \nabla^c h_{ab}
+ 4 \nabla_c \phi \nabla_{(b} h_{a)}^c \cr
& - {4\over (D-4)!}  h^{cd}
F_{aca_3...a_{D-2}} F_{bd}^{\;\;a_3...a_{D-2}} =0 & (2.39b) \cr
}
$$
the `electromagnetic' perturbation vanishing in this case. Replacing the
background values of the dilaton and Riemann tensor lead to the final versions
given in appendix B.
\def\half{{1\over 2}}

\proclaim 3) The zero charge case.

We will first give a detailed analysis for the case when the
black string or brane is uncharged.  This simplifies the calculation for two
reasons. The first is that the form of the  metric is much simpler, leading 
to more tractable coefficients in the perturbation equations of Appendix B.  
The second  is
that as can be seen from appendix B, some equations decouple from each other;  
in particular, the perturbation of the  dilaton can be neglected
and we can deal solely with the metric perturbations. In addition, as we have 
mentioned before, it is consistent to take a vanishing trace of the metric
perturbation. 

We can first investigate the transverse terms of the metric perturbations
in the $10-D$ dimensions.  These perturbations transform like scalars
under the $D-$dimensional coordinate transformations as is well known 
in Kaluza-Klein decompositions.  They are completely decoupled from all other
perturbations and obey equations of the form 
$$
A {h^{ii}}''  + B {h^{ii}}' + C h^{ii}=0
\eqno (3.1)
$$
where $A,B$ and $C$ are function of $r$.  Their explicit form can be
found in Appendix B.   We can investigate the behavior of $h^{ii}$ 
as $r\rightarrow \infty$ and and as $r\rightarrow r_+$. We obtain
$$ \eqalign{
h^{ii}&\sim \exp \pm r\sqrt{\Omega^2+\mu^2}\ \ \ r\rightarrow\infty  \cr
      &\sim  (r-r_+)^{\pm \Omega r_+}\ \ \     r\rightarrow r_+     \cr}
\eqno (3.2)
$$
The regular solutions correspond to the minus root at infinity
and the positive root at the horizon.  It is therefore easy to see
that a necessary condition to obtain a regular solution is to have a change in
sign in the ratio of coefficient $A/C$ in equation (3.1).  From 
appendix B it is seen that this ratio does not change sign for the transverse
 part of the metric perturbations.  We therefore set these
perturbations to zero in our search for an instability.

We can now study the vector perturbations $h^{\mu i}$.  An analysis
similar to the scalar case shows that they do not lead to an instability, they 
are therefore set to zero.  We are thus left with the tensor
perturbations.

The tensor perturbations are more complex as the equations are coupled with
each other.  It is possible to rewrite them in term of a single  variable,
$h^{tr}$ say, using the gauge conditions described in appendix B:
$$
\eqalign{
0&= \Bigl \{
\textstyle{- \Omega^2 - \mu^2 V +
  {(D-3)^2 \left ({r_+\over r} \right)^{2(D-3)}\over 4r^2} }\Bigr \}{H^{tr}}''\cr
&-\Bigl \{
\textstyle{ {\mu^2[(D-2)- 2 \left ({r_+\over r} \right)^{D-3} +
  (4 - D) \left ({r_+\over r} \right)^{2(D-3)}] \over rV } }\cr
&\textstyle{ +{\Omega^2[(D-2)+(2D-7)\left ( {r_+\over r} \right )^{D-3} ]
   \over rV }
- {3(D-3)^2\left ({r_+\over r} \right)^{2(D-3)}
  [(D-2)- \left ({r_+\over r} \right)^{D-3} ] \over
   4r^3V } }
\Bigr \} {H^{tr}}' \cr
&+ \Bigl \{ \textstyle{ \left( \mu^2 + \Omega^2 /V \right) ^2
+{\Omega^2 [ 4(D-2) - 8(D-2) \left ({r_+\over r} \right)^{D-3} -
       ( 53 - 34D + 5D^2 ) \left ({r_+\over r} \right)^{2(D-3)} ]
\over 4r^2V ^2} } \cr
& \textstyle{ +{\mu^2[ 4(D-2) - 4(3D-7)\left ({r_+\over r} \right)^{D-3} +
 ( D^2 +2D -11 ) \left ({r_+\over r} \right)^{2(D-3)} ]
 \over 4r^2V }}\cr
&\textstyle{ +{(D-3)^2 \left ({r_+\over r} \right)^{2(D-3)}
[(D-2)(2D-5) - (D-1)(D-2) \left ({r_+\over r} \right)^{D-3}
 + \left ({r_+\over r} \right)^{2(D-3)}] \over
4r^4 V^2}
} \Bigr\}  H^{tr}\cr
}
\eqno (3.3)
$$

Before dwelling on the existence of an instability for the tensor
perturbations we would like to comment on the flat space case, i.e $r_+=0$. 
In this case the ratio of the coefficients of the first and last
term does not change sign and we get the expected result that there is no
instability.

When $\Omega>(D-3)/r_+$ it is possible to use the same analytic argument 
as for the transverse perturbations to show that no instability exists. However
when $\Omega < (D-3)/r_+$
the ratio of the coefficients of the first and last term does indeed change sign
and we must resort to numerical techniques to establish the existence or 
otherwise of an instability. 

It is possible to reduce the set of second order equations to first order ones
by using the two gauge conditions (B.10a-b). By differentiating these and using
 the
second order equation we reduce them to first order equations but as we have
only three variables and four first order equations we can reduce one of them
to a constraint between the fields themselves (without derivatives).  One of the
variables can thus be reexpressed as a function of the others.

First let us define
$$\eqalignno{
H_\pm =& V_+ H^{tt} \pm {H^{rr}\over V_+}&(3.4a)  \cr
H     =& - H^{tr}                        &(3.4b)  \cr}
$$
where $V_+ = (1-(r_+/r)^{D-3})$.
The Lichnerowicz and the gauge equations reduces to the constraint:
$$\eqalign{
H_+ {V_+\over 2} \big [ \mu^2 + { (D-3)(D-2)\over 2r^2}&(1-V_+) \big ]
   = \cr
   &\ \  H_- \big [ \Omega^2 + {\mu^2 V_+ \over 2} + {(D-3)\over 4r^2}
              \big( -(D-3) + (D-4)V_+ + V^2_+\big) \big ]             \cr
   &   + H {V_+\over r} \big [  -\Omega(D-2) - {\mu^2\over 2 \Omega}
                                   \big ( (D-3) -(D-1)V_+\big) \big]  \cr}
\eqno (3.5)
$$
and the following two first order differential equations
$$
{\partial H \over \partial r} = {\Omega\over 2V_+} (H_+ + H_-)
    -{[(D-3) +V_+] \over rV_+} H
\eqno (3.6)
$$
and 
$$
{\partial H_- \over \partial r}=
 {\mu^2\over \Omega} H  + {(D-2)\over 2r} H_+
+ {[(D-3) + (-2D+3)V_+] \over 2rV_+} H_-
\eqno (3.7)
$$

>From this we can deduce that as $r\to \infty$
$$\eqalignno{
H   &= \sqrt{\Omega^2+\mu^2} F_+ e^{r\sqrt{\Omega^2+\mu^2}}
      -\sqrt{\Omega^2+\mu^2} F_- e^{-r\sqrt{\Omega^2+\mu^2}}  &(3.8a)    \cr
H_- &= {\mu^2\over \Omega} F_+ e^{r\sqrt{\Omega^2+\mu^2}}
      -{\mu^2\over \Omega} F_- e^{-r\sqrt{\Omega^2+\mu^2}}    &(3.8b)    \cr}
$$
and as  $r\rightarrow r_+$
$$\eqalignno{
H   =& (D-3) r^{D-4}_+ (-\half + {\Omega r_+\over D-3})G_+
              (r^{D-3} - r^{D-3}_+)^{-1 + {\Omega r_+\over D-3}}  \cr
     & -(D-3) r^{D-4}_+ (-\half - {\Omega r_+\over D-3})G_-
              (r^{D-3} - r^{D-3}_+)^{-1 - {\Omega r_+\over D-3}} &(3.9a) \cr
H_- =& ({\mu^2\over \Omega} + {D-2\over r_+} )G_+
              (r^{D-3} - r^{D-3}_+)^{{\Omega r_+\over D-3}}       \cr
     & -({\mu^2\over \Omega} - {D-2\over r_+} )G_-
              (r^{D-3} - r^{D-3}_+)^{- {\Omega r_+\over D-3}}.  &(3.9b)\cr}
$$
The behavior of $H_+$ can be obtained from the constraint (3.5).

The regular solution corresponds to $F_+ = 0$ and $G_-=0$.  We have used a
Runge-Kutta algorithm with variable stepsize to integrate these equation from
$r_1 =200$ to $r_2 =2.000002$ with a tolerance $\epsilon =10^{-6}$.  We do this
integration for fixed $\mu$ and various values of $\Omega$.  
Near $r=r_+$ we calculate the ratio $R=G_-/G_+$ and look for a change in
sign (as $\Omega$ varies).  When a change in sign occurs we home in towards
the value of $\Omega$ for which this change occurs and ensure that  the ratio
decreases towards such a value (an increase would imply that $G_-$ goes through 
a zero).  Figure 2 shows the behavior of the function $H^{tt},H^{tr}$ and $H^{rr}$.
 In Figure 3. we have plotted $\Omega$ as a function of $\mu$ for which
an unstable mode has been found. This plot takes $r_+=2$, but behavior
for other values of $r_+$ can be obtained from the scale transformation
$$
\eqalign{
r_+ &\rightarrow \alpha r_+ \cr
\Omega &\rightarrow{\Omega\over \alpha} \cr
\mu &\rightarrow{\mu\over \alpha} \cr}
\eqno (3.10)
$$

There are at least two interesting points in Fig.3.   The first one is to note
that $\Omega$ does not seem to go to zero as  $\mu\rightarrow 0$. This might
be a numerical artifact as our boundary conditions are different in this limit.
However if it is not a numerical artifact it would be surprising at first,
 as in this particular case the equations reduce exactly to
the stability of D-dimensional black holes.  We know that black holes in
four dimensions are stable,  moreover the mode under investigation here is the
spherically symmetric mode and we know by Birkhoff's theorem (and its
generalization to higher dimensions) that the solution must be 
Schwarzschild.  We
must therefore conclude that this mode is pure gauge.  We can in fact work out
what this gauge transformation looks like in the asymptotic regions.

As the system remains spherically symmetric and has no dependence on the extra
dimension, we must have 
$$
\xi^a = (\xi^t, \xi^r, 0,...,0)
\eqno (3.11)
$$
It is easy to verify that the particular gauge 
(in the large $r$ limit)
$$
\xi^t = \xi^r = {\rm e}^{\Omega(t-r)}
\eqno (3.12)
$$
corresponding to metric perturbations of the form:
$$
\eqalignno{
h^{tt} &= \xi^{t;t} \approx  -\Omega {\rm e}^{\Omega(t-r)} &(3.13a)  \cr
h^{rr} &= \xi^{r;r} \approx  -\Omega {\rm e}^{\Omega(t-r)} &(3.13b) \cr
h^{tr} &= \xi^{(t;r)} \approx  -\Omega {\rm e}^{\Omega(t-r)}. &(3.13c) \cr }
$$
This gives rise to a  transverse and traceless metric as can be easily verified.
In the $r\rightarrow r_+$ limit we have
$$
\xi^t = - r_+(r-r_+)^{-1+\Omega r_+}\ \ ; \ \ \xi^r =(r-r_+)^{\Omega r_+}
\eqno (3.14)
$$
corresponding to the metric perturbation
$$
\eqalignno{
h^{tt} & \approx   (-{1\over 2} + \Omega r_+)r_+ (r-r_+)^{-2+\Omega r_+} 
 e^{\Omega t}   &(3.15a) \cr 
h^{rr} & \approx {(-1 + 2\Omega r_+)\over 2r_+}(r-r_+)^{\Omega r_+}
 e^{\Omega t}   &(3.15b) \cr
h^{tr} & \approx  - (-{1\over 2} + \Omega r_+)
 (r-r_+)^{-1+\Omega r_+} e^{\Omega t}&(3.15c) \cr } 
$$
Of course we still have to show that there is a regular  solution which links
these two asymptotic behavior.  This can  be done  numerically.

The result for $\mu=0$ might leave the reader worried that the instability
found might be pure gauge.  However the case where $\mu\neq 0$ is different. In
this case it is easy to convince oneself that they cannot be pure gauge. The
simplest way is to assume we can write these perturbations in term of $\xi^a$
and find a contradiction.  Because perturbations in the $10-D$ dimensions are
zero, we would have $$ \xi^{j;j} = i\mu_j \xi^j = 0 \ \ \ j=D+1,...,10  
$$
thus this implies that $\xi^j = 0$ for $\mu_j\neq 0$ as $h^{jj} =0$.
If we now look at $h^{\mu j}=\xi^{\mu;j}$ this equation implies that 
$\xi^{\mu}=0$ and thus $h^{\mu\nu}=0$, a contradiction.  This ensures that the 
perturbations found in the uncharged case cannot be pure gauge and therefore
must be physical.  In the charged case the perturbations in the extra
dimensions are non-zero and thus such a proof is more difficult.  In this latter
case we will make use of gauge invariant variables described in appendix A to
show that the perturbations are not gauge artifacts.

It is interesting to understand where this extra degree of freedom for the
gravitational field has come from.  It is a result of the degree of freedom of
the gravitons due to the extra dimensions.  One easy way to make this
clear is to use a Kaluza-Klein analogy.  The gauge transformations in D
dimensions give the graviton degrees of freedom which are transverse to the
direction of propagation.  This is related to the masslessness of the graviton. 
Once the extra dimensions come into play, through the dependence $\exp i\mu_j
z^j$, this effectively gives a mass  to the graviton and thus a longitudinal
component corresponding to a spherically symmetric mode in D dimensions.  It is
this component which is responsible for the instability. In the limit as
$\mu\rightarrow 0$, this mode becomes pure gauge as the mass disappear.

The other interesting point to notice in Fig.3 is the fact that there is 
a maximum value $\mu_j^{max}$ for the instability to exist.  The 
analysis carried above
can be repeated in the case where the extra dimensions are 
periodically identified.  The only difference in this case is that  the values 
of 
$\mu_j$ will be quantized as $2\pi n/L_j$, for $n=0,1,...$  
and $L$ being the length of extra dimensions.  
>From this we can see that if the $L_j$ are small enough, the first 
value of $\mu_j$ will be larger than $\mu_j^{max}$ and thus
the instability will not occur.  The instability occurs only when the 
black holes (in D dimensions) have a Schwarzschild radius of the order of the 
size of the extra dimensions as suggested by the entropy argument 
described in the introduction.  Thus if we believe the assumptions of 
string theory that the $10-D$ extra dimensions are wrapped in small 
circles, the instability found 
here will not have any effect on astrophysical black holes.  It is interesting
that this instability appears to break the duality of the solutions obtained by replacing
the radius $L$ of the extra dimension by its inverse $1/L$.  If we  choose $L$
appropriately, there is an  instability for this spacetime but there will
not be one when $L\rightarrow 1/L$, however, as we have considered only momentum
modes in this calculation, and not winding modes, it is possible that there will be
a winding mode instability for the dual spacetime. However, we have neither
verified nor refuted this claim.

We have discussed the existence of the instability but what does it 
correspond to?   All the perturbations have the form $\exp i\mu_j z^j$
and thus implies an oscillatory behavior of the distance scale as a function
of the extra dimensions.  To make this clearer we can work out the location
of the apparent horizon.  To do so we introduce ingoing Eddington-Finkelstein
coordinates, and for ease of description restrict our attention to $D=4$: 
$$
ds^2  = -{(r-r_+)\over r}du^2  + 2dudr +r^2 d\Omega^2 +dz^2
\eqno (3.16)
$$
where $u= t+r + r_+ ln(r-r_+)$ in terms of the usual Schwarzschild coordinates.
The apparent horizon is located where the outgoing light-cones have zero
divergence.  The spherically symmetric light rays for  this metric
are given by 
$$
\cases{
u = {\rm const.} & ingoing \cr
{dr\over du} = {(r-r_+)\over 2r} & outgoing \cr
}
\eqno (3.17)
$$
In the case of the unperturbed black strings the apparent horizon is at 
$r=r_+$.   If we perturb the metric slightly we get for the new outgoing
null geodesics 
$$
h_{uu} - {(r-r_+)\over r} + 2{dr\over du}(1+h_{ur}) + h_{rr}
\left ({dr\over du}\right )^2 = 0
$$
which to first order gives
$$
{dr\over du} \approx  {(r-r_+)\over 2r} -{(r-r_+) \over 2r}h_{ur}  
      -{1\over 2}h_{uu} - {(r-r_+)^2\over 8r^2}h_{rr}
\eqno (3.18)
$$
which has a zero for 
$$
r\approx r_+ (1+ h_{uu}) + \lim_{r\to r_+}\left [ 
(r-r_+)h_{ur}+{(r-r_+)^2\over 4 r_+}h_{rr}\right ].
\eqno (3.19)
$$
Using the behavior of the solution obtained above we finally get
$$
r\approx r_+ + const. \cos(\mu z)
\eqno (3.20)
$$
This shows that the apparent horizon of the perturbed spacetime oscillates
as a function of the extra dimensions. The schematic behavior of the apparent
horizon is depicted in Figure 4. Thus it would appear that this perturbation
destabilises the event horizon causing it to ripple in the transverse
dimensions.

\proclaim  4) The small charge case.

We have seen in the  previous section that uncharged black strings and p-branes 
are unstable.  We now investigate the question whether a charge on these
black objects might stabilize them.  Physically, it seems reasonable that they 
will 
also be unstable at least for small charges, this is because for small charge
the metric is essentially the same as the uncharged case.  As long as we remain
outside the event horizon ($r>r_+$), as indeed we do to investigate the 
instability, the effect of the charge should be negligible. However a
non-trivial effect of the charge is that now the $10-D$ 
dimensional perturbations
that previously vanished become coupled to the D-dimensional non-zero
perturbations.  This now complicates the issue as we can no longer set
the former perturbations to zero; all the perturbations must
be solved simultaneously. An additional complication  is that it is not
possible to take  the trace $h=g_{ab}h^{ab}$ to be zero, for as we have
already shown, this simplification is only possible in vacuo.  The perturbation
of the scalar field $\phi$ couples to $h$ and prevents us from setting that part
of the metric to   zero.

In order to guide the reader through the maze of these equations we will
first focus on the small charge case.  Indeed in this case the problem,
although more complicated than the uncharged case, is drastically simplified
from the general charged case.   The simplification come through 
the observation that the coupling of D-dimensional perturbations 
to the $10-D$ ones
are always through a factor proportional to the charge $Q^2= r_-r_+/2$.
As the $10-D$ dimensional perturbations are vanishing for $Q=0$, we deduce
that they are in fact proportional to $Q$ at small charge.  This lead us to
an expansion of the perturbations in powers of $r_-$.  Using such an expansion 
 $h^{\mu\nu}$ will be zeroth order, and the other variables,
$h^{j\mu},h^{jk}$ and $\delta\phi$ will
be first order.  Notice that $h$, the trace of $h^{ab}$, will also be  first
order.

Before starting the analysis, note that from the form of the
perturbation equations in Appendix B, it is easy to see that writing
$$
\eqalignno{
h^{ij}&={i\mu_i\mu_j\over \mu^2}h^{zz} & (4.1a) \cr
h^{\mu i} &= {i\mu_i\over \mu} h^{\mu z} & (4.1b) \cr
}
$$
removes the $(10-D)$ $i$-degrees of freedom to just one transverse degree of
freedom, which we call `$z$'. Additionally, for calculational simplicity, we
will consider the specific case of the black 6-brane ($D=4$), 
which will at least
remove the variable $D$ from our equations! The analysis can be extended to
other black strings and p-branes, however, since we have already shown black
$p$-branes to be unstable for all values of $D$ in the previous section, an
analysis for general $D$ at this stage would be neither economical nor
illuminating.

We first analyze the equation for the perturbations of scalar field $f$,
$h^{zz}$ and the trace $h$:
$$\eqalign{
{(r - r_+ )\over r}f''
+{2r -r_+\over r^2}f' &
- \left( {\mu^2} + {\Omega^2 r  \over (r - r_+)}\right)  f
+ { r_-(r-2r_+)\over r^2} K  \cr
&+ {r_-(4r  -5r_+ ) \over 4r^3 (-r+r_+)} h^{rr}
- {r_-r_+(r-r_+) \over 4r^5} h^{tt} =0  \cr}
\eqno (4.2)
$$
$$
{(r - r_+ )\over r} {h^{zz}}''
+ {\left( 2r - r_+ \right)\over r^2} {h^{zz}}'
- \left( \mu^2 + {\Omega^2 r \over (r - r_+)} \right)  h^{zz}
+ 2 {\mu ^2} f(r) = 0
\eqno (4.3)
$$
$$
{(r - r_+ )\over r}({h''\over 2} -2f'')
+ {(2r - r_+)\over r^2}({h'\over 2} -2f')
- \left( \mu^2 + {\Omega^2 r \over (r - r_+)}\right)  ({h\over 2} -2f)
- {2r_-r_+ \over r^2} K   =0
\eqno (4.4)
$$
There is also an equation for $h^{rz}$ and $h^{tz}$ and the gauge condition
$$
\nabla_a {\bar h}^{az} \approx {h^{rz}}' + i \mu  h^{zz} + \Omega h^{tz}
 + {2\over r} h^{rz} - i{\mu\over 2} h \approx 0
\eqno (4.5)
$$
Taking a derivative of this gauge condition and using the $h^{tz}$ equation we
get $$
{(r-r_+)\over 2r} ( i\mu {h^{zz}}' - i{\mu\over 2} h' + \Omega {h^{tz}}'
+4i\mu f' )
+ \left( {\mu^2\over 2} + {\Omega^2 r\over 2(r-r_+)} \right ) h^{rz}
+ {i\mu r_- \over 2r^2} h^{rr}
+{\Omega r_+ \over 2r^2} h^{tz} \approx 0
\eqno (4.6)
$$
This gives us two first order equations for $h^{rz}$ and $h^{tz}$

With these equations we can investigate the behavior of the fields at infinity
and near $r=r_+$.  In the large $r$ limit all these fields have 
a behavior of the
form $e^{\pm r\sqrt{\Omega^2+\mu^2}}$, the regular solution having the negative
sign. The coefficients of $f,h, h^{zz}, h^{tz}$ 
are independent of each other but
once these are fixed, the coefficient in front of the exponential of $h_i^{rz}$
(denoted by  a subscript $i$)   is given by 
$$
h_i^{rz} \approx {1\over \sqrt{\Omega^2+\mu^2}}
             \left ( {\mu\over 2}h_i + i\Omega h_i^{tz} - \mu h_i^{zz} \right).
\eqno (4.7)
$$
in terms of the coefficients of the exponential of the other  perturbations.
Near $r=r_+$, the perturbations $h,f,h^{zz},h^{rz}$ behave as $(r-r_+)^{\pm
\Omega r_+}$ and $h^{tz}$ as $(r-r_+)^{-1 \pm \Omega r_+}$.

With these boundary conditions we can integrate eq(4.2-6) to find the unstable
mode assuming the zeroth order solutions found previously.  We have used a
similar  technique as in the uncharged case to find the  regular solution near
$r=r_+$.  By adjusting  the values at large $r$ we have been able 
to find regular solutions for all the perturbations. 
Fig.5 gives the behavior of the modes which lead to the instability.    

\proclaim 5) The charged case.

We have now shown that black strings and branes with
small  charges are unstable by assuming a perturbative analysis in term of the
charge. As we have seen in that case, we have a zeroth order equation in a small
charge  perturbation for the four dimensional metric perturbation.  The dilaton
and  the extra dimensions are first order  and could be solved assuming the
zeroth order solutions for the four dimensional metric.  In this section we give
the argument for the general charge.  The idea of how to handle the equations is
very much the same  as in the previous cases but the calculations are much more
tedious as the  equations will now involve the charge parameter (or rather $r_- =
2Q^2/r_+$).

The equations of motion for the  different fields are given in appendix B.
As before, we were unable to show the existence or absence
of an unstable mode analytically. We have thus resorted to numerical evaluation
of the unstable mode.  We can rewrite the set of equations in Appendix B
as a set of first order equations suitable for numerical integration.
First let us define a new variable $q=h/2-2f$ where $h$ is the trace of the 
metric perturbation and rewrite the equations for $f,q, h^{zz}$ as 
$$
\eqalign{
f'     = &\pi_f  \cr
\pi_f' = 
&-{r^2\over(r-r_-)(r-r_+)} \Big ( -(\mu^2 + \Omega^2{(r-r_-)\over (r-r_+)}) f
             + {r_-(r-r_+)(r_- - r_+) \over 4 r^3(r-r_-)^2}h^{tt}\cr
&   -{r_-(4r^2-5rr_- -5r_+ +6r_-r_+)\over 4r^2(r-r_-)^2(r-r_+)}h^{rr}\cr
         &   {r_-(r-2r_+)\over r^2}K + {(2r^2-rr_+ - r_-r_+)\over r^3} \pi_f       \Big ) \cr
q'     = &\pi_q  \cr
\pi_q' = & {r^2\over (r-r_-)(r-r_+)}
    \Big( {2r_-r_+\over r^2}K  + (\mu^2 + \Omega^2{(r-r_-)\over (r-r_+)}) q \cr
         &- {(r-r_-)(2r-r_+)\over r^3}\pi_q  \Big ) \cr
{h^{zz}}'     = &\pi_{h^{zz}}  \cr
\pi_{h^{zz}}' = & { r^2\over (r-r_-)(r-r_+)}
     \Big ( -4\mu^2 f + {i2\mu r_-\over r(r-r_-)}h^{rz} \cr
          &  (\mu^2 + \Omega^2{(r-r_-)\over (r-r_+)}) h^{zz} +
             {(- 2r + r_- + r_+)\over r^2} \pi_{h^{zz}} \Big ) \cr}            
\eqno (5.1)
$$

The next step is to use the gauge condition (B.10d), the gauge condition for the
index `$z$'. The equation for ${h^{tz}}'$
 is obtained by taking the derivative of this and using the equation of motion of
$h^{rz}$ to eliminate the second derivative.  This gives an equation in terms of
the variables and their first derivatives.  Explicitly we have 
$$ \eqalign{
i{h^{tz}}'= &  -{1\over \Omega} 
         \Big (  -{\Omega(r_--r_+)\over (r-r_-)(r-r_+)} ih^{tz}
                 -{\mu rr_-\over (r-r_-)^2(r-r_+)} h^{rr}     \cr
          & + \big( {\Omega^2 r^2\over (r-r_+)^2} 
            + {\mu^2 r^2\over (r-r_-)(r-r_+)}\big )ih^{rz} 
            -2r\mu\pi_f + \mu\pi_q - \mu \pi_{h^{zz}} \Big )   \cr
i{h^{rz}}' =& - \mu q  -2\mu f  -\Omega ih^{tz} 
             - {(2r-3r_-)\over r(r-r_-)} h^{rz}  +\mu h^{zz}   \cr}
\eqno (5.2)
$$
We can use the gauge condition for the indices $t$ and $r$ 
to get first order equations for $h^{tr}$ and $h^{rr}$
$$
\eqalign{
{h^{tr}}'= & - {\Omega(r-r_-)(q+2f)\over 2(r-r_+)} -\Omega h^{tt}
            - {(2r^2 -4rr_- -rr_+ +3r_-r_+)\over r(r-r_-)(r-r_+)}h^{tr}
            -i\mu h^{tz}  \cr
{h^{rr}}'= &   {(r-r_-)(r-r_+)\over r^3}(2q+4f) 
           - { (r-r_+)(-2r^2 + rr_- +3rr_+ -2r_-r_+)\over 2r^3(r-r_-)}h^{tt} \cr
           & -\Omega h^{tr} 
    - {(6r^2-9rr_- - 7rr_+ + 10r_-r_+)\over 2r(r-r_-)(r-r_+)} h^{rr} - i\mu
h^{rz} \cr
           & - {(r-r_-)(r-r_+)\over r^3} h^{zz} 
             + {(r-r_-)(r-r_+)\over r^2}(\pi_q+2\pi_f)   \cr}
\eqno (5.3)
$$
Taking a derivative of the gauge condition  with index $t$ and getting rid of 
the second order derivative using the second order equation for $h^{tr}$ we
get 
$$ \eqalign{
{h^{tt}}'= & -{1\over \Omega}  \Big (
               - {\Omega (r_- - r_+)\over (r-r_+)^2}q  
               - {\Omega (r_- - r_+)\over (r-r_-)(r-r_+)}h^{tt} + \cr
 &  \big ({\Omega^2 r^2\over (r-r_+)^2} + {\mu^2r^2\over (r-r_-)(r-r_+)} \big )
h^{tr}  + {i\mu(r_- - r_+)\over(r-r_-)(r-r_+)} h^{tz}  +\cr
 &  \big  ( {\mu^2r r_-\over \Omega(r-r_-)^2(r-r_+)} +
{\Omega r (-2rr_- + rr_+ + r_-r_+)\over (r-r_-)(r-r_+)^3} \big )h^{rr} \cr 
           & - {i\mu\over \Omega}
\big( {\Omega^2r^2\over(r-r_+)^2} +{\mu^2r^2\over (r-r_-)(r-r_+)} \big )
h^{rz}    - \big( {\mu^2\over \Omega} - {\Omega(r-r_-)\over (r-r_+)}\big )
(\pi_q -2\pi_f) \cr
 &   +{\mu^2\over \Omega}\pi_{h^{zz}}   \Big )  \cr}
\eqno (5.4)
$$

We have now the field equations written in terms of first derivatives but 
we still have an equation left which comes from the time derivative of the
gauge condition with index $r$.  This leaves us with a constraint between the
variables in equations (5.1-4) given by 
$$ \eqalign{
&\big( {2\Omega^2(r-r-_)^2\over r^2} + {2\mu^2(r-r_-)(r-r_+)\over r^2}
       +{(r_- - r_+)(-2r^2+rr_- +3rr_+ - 2r_-r_+)\over 2r^5} \big) f \cr
&+\big(  {(4r^3r_- - 3r^2 r_-^2 -10r^2r_-r_+ + 8rr_-^2r_+ + r^2r_+^2 +
         4rr_-r_+^2 -4r_-^2r_+^2)\over 8r^6} \cr
 &\ \ \
        - {\mu^2 (r-r_-)(r-r_+)\over 2r^2} - {\Omega^2(r-r_-)^2\over 2r^2}
  \big) h  \cr
&+\big( {(r-r_+)(-r^2r_-^2 + 4r^3r_+ - 8r^2r_-r_+ + 6rr_-^2r_+ - 3r^2r_+^2 
         + 6rr_-r_+^2 -4r_-^2r_+^2)\over 8r^6(r-r_-)} \cr
  &\ \ \  - {\Omega^2(r-r_-)(r-r_+)\over 2r^2}    \big )h^{tt}  \cr
&+\big( {\mu^2(r-r_+)(2r^2-rr_- - 3rr_+ + 2r_-r_+)\over 4\Omega r^3(r-r_-)} +
        {\Omega(2r-r_-)(r-r_+)\over 2r^3} \big ) h^{tr}  \cr
&+\big(-{i\Omega\mu(r-r_-)(r-r_+)\over r^2} +
      {i\mu(r_--r_+)(r-r_+)(2r^2-rr_- -3rr_+ +2r_-r_+)\over 4\Omega r^5(r-r_-)}
   \big) h^{tz}   \cr
&+\big(         {(-16r^3r_- +11r^2r_-^2 + 34 r^2r_-r_+ -24rr_-^2r_+ - r^2r_+^2
          -16rr_-r_+^2  +12r_-^2r_+^2 \over 8r^4(r-r_-)(r-r_+)}\cr
  &\ \ \  -{\mu^2r_-(r-r_+)(-2r^2+rr_- + 3rr_+ -2r_-r_+)\over 4\Omega^2r^4(r-r_-)^2}
+{\mu^2\over 2} + {\Omega^2(r-r_-)\over 2(r-r_+)} \big) h^{rr} \cr
&+\big( {i\mu(-4r+3r_-)(r-r_+)\over2r^3}
      +{i\mu^3(r-r_+)(2r^2-rr_- -3rr_+ + 2r_-r_+)\over 4\Omega^2r^3(r-r_-)}
   \big)h^{rz} \cr
&+\big( {(r-r_-)(r-r_+)(-rr_- -rr_+ +2r_-r_+)\over 4r^6}
      +{\mu^2(r-r_-)(r-r_+)\over 2r^2}  \big) h^{zz}  \cr
&+\big( {2(r-r_-)(-3r+2r_-)(r-r_+)^2\over r^5}
      +{\mu^2(r-r_+)^2(2r^2-rr_- -3rr_+ + 2r_-r_+)\over \Omega^2 r^5}
   \big)\pi_f   \cr
&+ {\mu^2(r-r_+)^2(-2r^2 +rr_- +3rr_+ -2r_-r_+)\over 8\Omega^2 r^5} \pi_h  \cr
&+\big( {(r-r_-)^2(r-r_+)^2\over 2r^5}  +
        {\mu^2(r-r_+)^2(2r^2 - rr_- -3rr_+ + 2r_-r_+)\over 4\Omega^2r^5}  
   \big) \pi_{h^{zz}} = 0 \cr }
\eqno (5.5)
$$
The initial data (as $r\rightarrow\infty$) must obey this constraint, and then is 
automatically preserved along the integration.  
We have used it to verify that the integration errors were indeed small.

In the previous sections we 
deduced the existence of an instability by finding zeros of a ratio, $R$, of
the irregular to regular solutions near the horizon. As it happened, the
functions were essentially independent near the horizon, so we could
investigate their ratios independently. Changing the dilaton perturbation,
say, and thus $R_{dil}$, did not affect the ratios for the other perturbations,
therefore we could vary our parameters freely to find the instability. However,
when the charge becomes large, this is no longer the case, changing the initial
value for the dilaton perturbation will now change the ratio of field $h_{zz}$,
say. The task of finding where a 5-dimensional vector (corresponding to the
ratio of each of the perturbation fields) is much harder. We have thus resorted
to the Newton-Raphson method described in [17]. In a nutshell the
idea is to assume that we are not too far from the zero 
$$ 
0= R_i (x +\delta x)\approx  R_i(x) + {\partial R_i\over \partial x_j}\delta x_j
\eqno (5.6) 
$$ 
We can then invert this equation to get  
$$ 
\delta x_j \approx
{\partial R_i\over \partial x_j}^{-1} R_i(x) 
\eqno (5.7) 
$$ 
that is we calculate
the gradient of $R$ at the value of x we are and follow  it in towards the zero
of  $R$.

Figure 5 shows $\Omega$ for an unstable mode as a function of the charge for 
different values of $\mu$, the frequency in the orthogonal direction.  The
important point to note is that the curves cross zero, at which point the
unstable mode disappears for the appropriate value of $\mu$.

We have checked using the gauge invariant variables of appendix A
that the solutions found correpsond to real physical solution and not
gauge artifacts.  This is obvious for the small charge case as the dominant
quantities entering in equations (A.4) and (A.5) are the chargeless terms
with the small charge terms being sub-dominant. For large charges  
we have calculated the quantity (A.4) directly. It is non-zero for 
the perturbations studied.

Near the extreme caase, the function calculating the ratio between the
regular and irregular  modes has very large derivatives and thus is
not
easily amenable to numerical study. As this happens to be also
at very small values of $\Omega$ we conjecture that the extreme black hole is 
stable.  This important particular case will be investigated elsewhere
[18].

\proclaim 6) Conclusion.

We have investigated the instability of `magnetically' charged dilatonic black
$p$-branes in  string theory. In particular we have  shown that the
chargeless instability found in [13] still persists in the presence of a charge
on the black string or brane.  This demonstrates that this instability is a
generic phenomena.  We remind the reader that this instability can be pictured
as the string horizon collapsing in some regions of the extra dimensions and
expanding in others.  The unstable mode remains present for large values of
charge but in this paper we have not proved its existence or otherwise for the
extremal case;  from the trend of the time frequency $\Omega$ as function
of the charge it seems rather plausible that this case might be stable.
However, without a more detailed analysis tailored to the extremal case, we
cannot be definite. We can however indicate a different way of stabilizing
the black $p$-brane - compactification. Compactification  implies that the
values of $\mu_i$ are quantized. If the compactification is on a scale smaller
than the inverse mass of the black hole, the first allowable value of $\mu$
would be greater than the maximum one allowed for the instability, so such black
doughnuts would be stable. Since there must be compactification of any extra
dimensions on an extremely small scale, all but the tiniest black doughnuts
would be safe, and those that would not would presumably have evaporated
producing their own naked singularities long ago.  Thus this instability will 
have no effect for contemporary astrophysical black holes.

One of the main questions of interest about this instability is the nature of
its endpoint - what does the black string become? Of course, since our
calculation is linear, we cannot strictly say anything about the final state,
however, since in the chargeless case there are no
other scales in the problem except the mass per unit length of the black
string, it seems very reasonable to conjecture that the endpoint of the
instability will be the fragmentation of the black strings into a bunch of small
spherical black holes.  This suggestion is supported by the entropy argument
given in the first section. Periodic black hole solutions are known$^{19}$, so
unlike the Reissner-Nordstrom instability, there is a final state solution
here. It is quite likely that such a solution will itself exhibit a Jeans-like
instability to long wavelength clustering of black holes which itself will be
unstable and so on. In any case, a process of fragmentation will produce a
naked singularity and hence violate cosmic censorship, however, we should again
stress that this is speculative; ideally one should follow the instability
numerically into the non-linear r\'egime to see if fragmentation develops, this
work is currently underway.

For the charged black $p$-brane, there are other possibilities. The fact that
$\delta F=0$ suggests that where the horizon shrinks the solution becomes
locally closer to extremality. It is possible that when a charge is
present on the black string or brane fragmentation will
not occur.  The electromagnetic repulsion might stabilize the string as the
2-spheres where the charge resides shrink to smaller volume.  Unlike the
chargeless case, the charged case might not lead to  naked singularities and
fragmentation. This conclusion assumes both the stability of extremal black
$p$-branes, which is likely, as well as the non-participation of charge in the
instability, which is an unknown without following the instability into the
non-linear r\'egime. Alternatively, the `magnetic' nature of the charge
considered also makes it plausible that as the horizon shrinks in some regions,
higher energy physics might come into play, and that in such regions, the
relevant monopole solution could `pop out' from behind the horizon. For
example, consider $D=4$. Here an instability is known for the four dimensional
monopole black hole, in which for small enough horizons, the t'Hooft Polyakov
monopole is the preferred exterior solution$^{20}$. Thus, in the case of the
6-brane should the horizon volume  become too small,  there will be sections of
the rippled brane that will have t'Hooft Polyakov type monopole fields
surrounding them. The charge, having appeared in an exterior smooth form,
would then presumably pose no obstruction to the interior collapse of the
event horizon, and thence a naked singularity. Again, this assumes some
knowledge of the non-linear r\'egime. 

We can also comment on black $p$-branes with topological charge. For example,
consider the axionic black holes of Bowick et.~al.$^{21}$ These
carry `quantum' charge, detectable only by an Aharanov-Bohm scattering process,
but as far as classical physics is concerned (hence our instability) they
behave as if uncharged, since the quantum charge they carry does not affect the
exterior spacetime which is consequently Schwarzschild. The field strength of the
axion field is zero throughout the spacetime, but the gauge field (rather like
the gauge field of a local cosmic string) is non-trivial due to the topology of
the spacetime, $B = {Q\over \sin\theta }d\theta \wedge d\phi$. This solution can
clearly be extended to a five dimensional black string (or ten-dimensional black
6-brane). This solution too will be unstable, however, a five (or
ten) dimensional black hole cannot carry the same type of axion charge. Drawing
analogy to the gauge field of a local cosmic string, it seems likely that during
the fragmentation process, higher energy physics could come into play, producing
an axion vortex, which could appear from behind the event horizon by an analogous
process discussed above for the charged black $p$-brane. The endpoint in this
case might  be a line of black holes threaded by a cosmic axion string (not
to be confused with the four-dimensional global string).

To summarize, whether charged or uncharged, black $p$-branes exhibit a long
wavelength instability which causes the apparent horizon to ripple. A plausible
endpoint of this instability is fragmentation, and hence violation of cosmic
censorship, however, without more evidence in the non-linear r\'egime, this
remains speculative. Perhaps a more realistic conclusion is that due to this
instability, black strings and $p$-branes will not form from collapse in the
first place. From our work, it appears there are two ways of avoiding this
instability. One is if the black brane is extremal, although the methods
described here were unable to probe this end of parameter space, all
indications were that these solutions were stable. This is also anticipated
since extremal branes are supersymmetric. The other way of avoiding the
instability is by compactification, indeed, to be totally speculative, since
the endpoint of the instability does appear to be some periodic solution, it is
tempting to suggest that the instability of some primordial black string could
have triggered an effective compactification of our universe!
  
\proclaim Acknowledgments.

We are grateful to J.B.Hartle, J.A.Harvey, S.W.Hawking, G.Horowitz, D. Wiltshire, 
and  W.H.Zurek for useful conversations. We appreciate the
careful perusal of the manuscript and accompanying remarks by
J.Bowcock and P.Laflamme. R.G. is supported by the
S.E.R.C. R.L. thanks Los Alamos National Laboratory for support.
R.G. also acknowledges the support of the McCormick Fellowship at the Enrico Fermi
Institute where this work was started.

\parskip=0pt
\newcount\refno
\refno=0
\def\nref#1\par{\advance\refno by1\item{[\the\refno]~}#1}

\def\book#1[[#2]]{{\it#1\/} (#2).}

\def\annph#1 #2 #3.{{\it Ann.\ Phys.\ (N.\thinspace Y.) \bf#1} #2 (#3).}
\def\cmp#1 #2 #3.{{\it Commun.\ Math.\ Phys.\ \bf#1} #2 (#3).}
\def\mpla#1 #2 #3.{{\it Mod.\ Phys.\ Lett.\ \rm A\bf#1} #2 (#3).}
\def\ncim#1 #2 #3.{{\it Nuovo Cim.\ \bf#1\/} #2 (#3).}
\def\npb#1 #2 #3.{{\it Nucl.\ Phys.\ \rm B\bf#1} #2 (#3).}
\def\plb#1 #2 #3.{{\it Phys.\ Lett.\ \bf#1\/}B #2 (#3).}
\def\prd#1 #2 #3.{{\it Phys.\ Rev.\ \rm D\bf#1} #2 (#3).}
\def\prl#1 #2 #3.{{\it Phys.\ Rev.\ Lett.\ \bf#1} #2 (#3).}

\proclaim References.

\nref
S.W.Hawking, \cmp 43 199 1975.

\nref
S.W.Hawking, \cmp 87 395 1982.

\nref
E.Poisson and W. Israel, \prd 41 1796 1990.

\nref
G.Gibbons and K.Maeda, \npb 298 741 1988.

D.Garfinkle, G.Horowitz and A.Strominger, \prd 43 3140 1991.

\nref
R.Gregory and J.Harvey, \prd 47 2411 1993.

J.Horne and G.Horowitz, \npb 399 169 1993.

\nref
R.Myers and M.J.Perry, \annph 172 304 1986.

\nref
G.T.Horowitz and  A.Strominger, \npb 360 197 1991.

\nref
W.Israel, \cmp 8 245 1968.

S.W.Hawking, \cmp 25 152 1972.

\nref
R.Gregory and R.Laflamme, \prd 37 305 1988.

\nref
C.V. Vishveshwara, \prd 1 2870 1970.

\nref
P.Bizon and R.Wald, \plb 267 173 1991.

\nref
B. Whitt, Ph.D Thesis, Cambridge, 1988.

\nref
R.Gregory and R.Laflamme, \prl 70 2837 1993.

\nref
S.W.Hawking and G.F.R.Ellis, \book The Large Scale Structure of Spacetime
[[Cambridge University Press 1973]]

\nref
R.M.Wald
\book General Relativity [[Chicago University Press]]

\nref
J.Bardeen, \prd 22 1882 1980.

\nref
W.H.Press, B.P.Flemming, S.A.Teukolsky and W.T. Vettering,
\book Numerical Receipes [[Cambridge University Press 1988]]

\nref
R.Gregory and R.Laflamme, The stability analysis of the extremally
charged black string, in preparation.

\nref
A.Bogojevic and L.Perivolaropoulos, \mpla 6 369 1991.

\nref
K.Lee and E.Weinberg, \prl 68 1100 1992.

\nref
M.Bowick, S.Giddings, J.Harvey, G.Horowitz and A.Strominger, \prl 61 2823 1988.

\proclaim Appendix A.  Gauge invariant variables.

One of the worries of results in linear perturbations of a system with gauge
invariance is to insure that the result are not artifact of the particular
gauge chosen.  One easy way to deal with this problem is to use gauge
invariant variables.  In this appendix we explain how to construct these 
variables.  

Under a coordinate transformation
$$
x^a \rightarrow x^a +\xi^a
\eqno (A.1)
$$
we have the following transformation of the fields
$$
\eqalign{
h^{ab}&\rightarrow h^{ab} + \xi^{(a;b)}   \cr
\phi  &\rightarrow \phi + \xi^a D_a \phi  \cr
\delta F_{a_1...a_{D-2}} &=  F_{a_1...a_{D-2};d} \xi^d + 
(D-2) F_{d[a_2...a_{D-2}} \xi^d_{;a_1]} \cr}
\eqno (A.2)
$$
In the metric given in eq. (2.3-2.5), they have the explicit form
(for $V_+=(r-r_+)$ and $V_-=(r-r_-)$)
$$
\eqalignno{
h^{tt} &\rightarrow h^{tt} - {V_-\over V_+}\Omega \xi^t
                     -  {(r_=-r_-)\over 2V_+^2} \xi^r  & (A.3a) \cr
h^{rr} &\rightarrow h^{rr} + {V_- V_+\over^2}{\xi^r}'  
                     + {(-rr_- - rr_+ + 2r_+r_-)\over 2r^3}\xi^r & (A.3b)  \cr
h^{rt} &\rightarrow h^{rt} + {V_-V_+\over 2r^2}{\xi^r}'
                                  -{V_-\over 2V_+}\Omega\xi^r & (A.3c) \cr
h^{\theta\theta} &\rightarrow h^{\theta\theta} + {\xi^r\over r^3} &(A.3d) \cr
h^{zt} &\rightarrow h^{zt} - {V_-\over 2V_+}\Omega \xi^z 
                          + {i\mu\over 2}\xi^t  & (A.3e)   \cr 
h^{zr} &\rightarrow h^{zr} + {V_-V_+\over 2r^2}{\xi^z}' 
                           + {i\mu\over 2}\xi^r & (A.3f) \cr 
h^{zz} &\rightarrow h^{zz} + {i\mu\over 2}\xi^z & (A.3g)\cr }
$$ 

It is easy to show that the linear combination
$$
Y= h^{zr} + i{V_-V_+\over 2\mu r^2}{h^{zz}}' - {i\mu r^3\over 2}
h^{\theta\theta}
\eqno (A.4)
$$
is indeed invariant under gauge transformation.  However this gauge invariant
variable is not well defined when $\mu =0$, a case under consideration
in section 3.  A more complicated variable
$$
\eqalign{
X= & h^{tr} + {(r_+-R_-)V_+ \over 2 r^2V_-\Omega}h^{tt}
     + {V_+^2\over 2r^2\Omega}{h^{tt}}'                              \cr
  & - {(r_+-r_-)(-r^2+2rr_+ + 2rr_- - 3r_+r_-)\over V_+V_-\Omega}K
    +{(r_+-R_-)r\over 4\Omega}K'
    -{V_-\Omega r^3\over 2V_+}K                                       \cr}
\eqno (A.5)
$$
can be used to investigate the $\mu\rightarrow 0$ limit in the chargeless
case.   In this limit we get $X\propto \mu$ as can be shown from the
asymptotic region  ($r\rightarrow\infty$ and $r\rightarrow r_+$) discussed in
section 3.  This analysis shows that the perturbations described
in sections 3-5 describe real physical instabilities except in the case 
$\mu=0$.


\overfullrule=0pt       
\def\bo{ { \sqcup\llap{ $\sqcap$} } }

\proclaim Appendix B)  Background Quantities.

The background metric is given explicitly by the following:
$$\eqalign{
g_{tt} =& - { r^{D-3}-r_+^{D-3} \over r^{D-3}-r_-^{D-3} }      \cr
g_{rr} =& { r^{2(D-3)} \over (r^{D-3}-r_-^{D-3})(r^{D-3}-r_+^{D-3})}  \cr 
g_{\theta_1\theta_1} =& r^2                \cr
g_{\theta_\alpha\theta_\beta} =&  r^2\delta_{\alpha\beta}
\prod_{n=1}^\alpha\sin ^2\theta_n  \;\;\;{\rm for}\;\alpha>1   \cr
g_{ij} =& \delta_{ij}                           \cr}
\eqno (B.1)
$$
where $\alpha, \beta$ run from 1 to $D-2$, and represent the angular
variables. The `matter' content has solution:
$$
\Phi = -{1\over 2} \ln \left \{ 1-\left ({r_-\over r}\right)^{D-3} \right \}
\eqno (B.2)
$$
$$
F = Q \epsilon_{D-2}\; , \;\;\; Q^2 = {(D-3)\over 2} (r_+r_-)^{D-3}
\eqno (B.3)
$$
The form of the perturbation is
$$
g^{ac}g^{bd}\delta g_{cd} = 
h^{ab} = e^{\Omega t+i\mu_ix^i} \left [
\matrix{ H^{ij}(r)& H^{it} &H^{ir}& 0  &0&... \cr
         H^{tj} & H^{tt} & H^{tr} & 0 &0&...\cr
         H^{rj} & H^{tr} & H^{rr} & 0 &0&...\cr
		       0 & 0 & 0 & K(r)&0&...\cr
	        0 & 0 & 0 & 0 &K/\sin^2\theta &...\cr
          ...&...&...&...&...&...\cr} \right ]
\eqno (B.4)
$$
$$
\delta \Phi = e^{\Omega t+i\mu_ix^i}  f(r)
\eqno (B.5)
$$
and finally
$$
\delta F = 0
\eqno (B.6)
$$

The non-zero elements of the Riemann tensor are given by:

$$\eqalignno{
R^{t}_{\ rtr}=& {(D-3)\over 2r^2} 
{\Bigl [ ({r_+\over r})^{D-3} - ({r_-\over r})^{D-3} \Bigr ]
\Bigl[ D-2-({r_-\over r})^{D-3} \Bigr ]
\over \Bigl [ 1 - ({r_+\over r})^{D-3} \Bigr ]
\Bigl [ 1 - ({r_-\over r})^{D-3} \Bigr ] ^2 }
& (B.7a)\cr
R^{t}_{\ \theta_alpha t\theta_\beta}=&{-(D-3)\over2r^2} \Bigl [ 
({r_+\over r})^{D-3} - ({r_-\over r})^{D-3} \Bigr ] 
g_{\alpha\beta}& (B.7b)\cr
R^{\theta_\alpha}_{\ r\theta_\beta r}=& {-(D-3)\over 2r^2}
{\Bigl [({r_+\over r})^{D-3} + ({r_-\over r})^{D-3}- 2({r_+r_-\over r^2})^{D-3}
\Bigr ] \over \Bigl [ 1 - ({r_+\over r})^{D-3} \Bigr ]
\Bigl [ 1 - ({r_-\over r})^{D-3} \Bigr ] } 
\delta^\alpha_\beta& (B.7c)\cr
R^{\theta_\alpha}_{\ \theta_\beta\theta_\gamma\theta_\delta}
=&{1\over r^2} \left [ ({r_+\over r})^{D-3} 
+({r_-\over r})^{D-3}-({r_+r_-\over r^2})^{D-3} \right ]
(\delta^\alpha_\gamma g_{\beta\delta} - \delta^\alpha_\delta
g_{\beta\gamma}) & (B.7d)\cr}
$$

Ricci tensor

$$\eqalignno{
R_{tt} =& -{(D-3)^2 \over 2r^2} \left ({r_-\over r} \right )^{D-3}
{\Bigl [ ({r_+\over r})^{D-3}  - ({r_-\over r})^{D-3}\Bigr ] 
\Bigl [ 1 - ({r_+\over r})^{D-3} \Bigr ] \over \Bigl [
1 - ({r_-\over r})^{D-3} \Bigr ] ^2 }  & (B.8a)\cr
R_{rr} =& {(D-3)({r_-\over r})^{D-3}
\left [ (3D-7) ({r_+\over r})^{D-3} + (D-1) ({r_-\over r})^{D-3}
- 2(D-2)\left (1+({r_+r_-\over r^2})^{D-3}\right ) \right ] 
\over 2r^2 [1 - ({r_-\over r})^{D-3}]^2[1-({r_+\over r})^{D-3}]}& \cr
&& (B.8b) \cr
R_{\theta_\alpha\theta_\beta}=& {(D-3)\over r^2}
\left ({ r_-}\over r\right )^{D-3} g_{\alpha\beta} & (B.8c)\cr}
$$

Ricci scalar
$$
R={(D-3)({r_-\over r})^{D-3}-\over r^2(1-({r_-\over r})^{D-3})}
 [-(D-4)-(D-2)({r_+\over r})^{D-3}({r_-\over r})^{D-3} 
+(2D-5)({r_+\over r})^{D-3} - ({r_-\over r})^{D-3}]
\eqno(B.9)
$$

The gauge conditions component by component are given by
the transverse conditions $\nabla_a \bar h^{ab} =0$. For
simplicity in what follows, we will write $R_\pm=(r_\pm/r)^{D-3}$
$$\eqalignno{
\nabla_a \bar h^{at} =&\
  {\Omega h^{tt}\over 2}  + 
  {[ (D-2) - (3D-8)R_- -  R_+ + (2D-5)R_-R_+ ]\over 
    {r (1-R_+)(1-R_-)}} h^{tr} + 
  i \mu_i  h^{ti}(r) +  
  {\Omega\over 2(1-R_+)^2} h^{rr} & \cr
& +
  {(D-2)\Omega r^2(1-R_-)\over 2(1-R_+)} K +
  {\Omega  (1-R_-)\over 2 (1-R_+)} h^i_i +
  {h^{tr}}'(r)   =0  & (B.10a)\cr
\nabla_a \bar h^{ar} =&\
  {  (D-3)(1-R_+)  (R_+ - R_-)\over r  (1 - R_-) }h^{tt}  
  -{[(D-2)-(2D-5)R_-]\over r(1 - R_-)} h^{rr} \cr
& -
  2(D-2)r(1-R_-)(1-R_+) K 
   + \Omega   h^{tr} 
+  i \mu_i   h^{ri} 
+ 
  {(1-R_+)^2\over 2} {h^{tt}}' \cr
&+{{h^{rr}}' \over 2} -
   {(D-2)\over 2}(1-R_-)(1-R_+)r^2 K' -
  {(1-R_-)(1-R_+) \over 2} {h^i_i}'
   =0 & (B.10b)\cr 
\nabla_a \bar h^{a\theta} =&\ \nabla_a \bar h^{a\phi} = 0 &(B.10c)\cr
\nabla_a \bar h^{ai} =&\
  {i \mu_i (1-R_+) \over 2 (1-R_-)}h^{tt} +
  \Omega   h^{ti} -
  {i \mu_i \over 2  (1-R_+) (1-R_-)}h^{rr} + 
  {[(D-2) - (2D-5)R_-] \over r(1-R_-)}  h^{ri}\cr
& -
  {(D-2)\over2}i \mu_i r^2 K 
+ i\mu_jh^{ij} -
  {i\mu_i\over 2} h^j_j +  
  {h^{ri}}'(r)  =0   & \cr
&  &(B.10d)\cr}
$$

Finally we give equations for the perturbations $h^{ab}$, the trace
$h=g_{ab}h^{ab}$ and $f=\delta\Phi$.

$\bullet h^{tt}$
$$
0= {-(D-3)h^{rr}\over 2 r^2}{(R_+-R_-)
[2(D-2)-(D-1)R_+-(3D-7)R_-+2(D-2)R_+R_-]\over(1-R_+)^3(1-R_-)}
$$
$$
+ h^{tt} \Bigl ( -\mu^2 - {\Omega^2(1-R_-)\over(1-R_+)}
+ {(D-3)^2\over2}{(R_+-R_-)^2\over(1-R_+)(1-R_-)} 
\Bigr )
+ (D-2) (D-3)K {(R_+-R_-)(1-R_-)\over(1-R_+)}
$$
$$
-  {{ 2(D-3)\Omega  }\over r} {(R_++R_+R_--2R_-)\over(1-R_+)^2} h^{tr}
-4\Omega^2{(1-R_-)^2\over(1-R_+)^2}f(r)
+ {2(D-3)\over r} {(1-R_-)(R_+-R_-)\over(1-R_+)}  f'
$$
$$
+ [(D-2)-(2D-5)R_-+(2D-7)R_+-(D-4)R_+R_-]  {{h^{tt}}'\over r}
+ (1-R_+)(1-R_-) {h^{tt}}''
\eqno (B.11a)
$$

$\bullet h^{rr}$
$$
0= {-(D-3)h^{tt}\over 2 r^2(1-R_-)}(1-R_+)(R_+-R_-)
[2(D-2)-(D-1)(R_++R_-)+2R_+R_-]
$$
$$\eqalign{
&+ h^{rr} \Bigl ( -\mu^2 - {\Omega^2(1-R_-)\over(1-R_+)}
-{4(D-2)(1-2R_+-(D-1)R_-)\over 2r^2(1-R_+)(1-R_-)} \cr
&+ {{(3D^2-26D+47)R_-^2+(D^2-10D+17)R_+^2 +2(2D^2-14D+23)R_+^2R_-^2 }\over
2r^2(1-R_+)(1-R_-)} \cr
&-{  2(D^2-5)R_+R_- 
+(D^2-12D+23)R_+^2R_- + (5D-13)(D-5)R_-^2R_+ \over r^2(1-R_+)(1-R_-)} 
\Bigr )
\cr}
$$
$$
+ (D-2) K (1-R_+)(1-R_-)[2+2(D-2)R_+R_- - (D-1)(R_++R_-)]
$$
$$
-  {{ 2(D-3)\Omega ( R_+-R_-) }\over r}  h^{tr}
+ [(D-2)-(2D-5)R_+-(4D-11)R_-+(5D-14)R_+R_-]  {{h^{rr}}'\over r}
$$
$$
-  {2(D-3)\over r} (1-R_+)(1-R_-)[R_++R_--2R_+R_-]  f'
-4(1-R_+)^2(1-R_-)^2f''(r)
+ (1-R_+)(1-R_-) {h^{rr}}''
\eqno (B.11b)
$$

$\bullet h^{rt}$
$$
0={\Omega(D-3)\over r} \left ( h^{rr}{[2R_--R_+-R_+R_-]\over(1-R_+)^2}
- h^{tt}(R_+-R_-) \right ) - 2\Omega f{(D-3)(R_+-R_-)(1-R_-)\over r(1-R_+)}
$$
$$+ h^{rt} \Biggl ( -\mu^2 - {\Omega^2(1-R_-)\over(1-R_+)}
- {{(D-2)[1-2(D-2)R_- +(D-5)R_+]\over r^2(1-R_+)(1-R_-)}}
$$
$$
+{(D^2-10D+20)R_-^2-R_+^2 - (2D-5))R_+^2R_-^2 \over
r^2(1-R_+)(1-R_-)} 
$$$$
+{(3D^2-17D+26)R_+^2R_- - (3D^2-13D+16)R_+R_- 
- (2D^2-18D+34)R_-^2R_+ \over r^2(1-R_+)(1-R_-)} 
\Biggr )
$$
$$
+4\Omega(1-R_-)^2 f' 
+ [(D-2)-R_+-(3D-8)R_-+(2D-5)R_+R_-]  {{h^{rt}}'\over r}
+ (1-R_+)(1-R_-) {h^{rt}}''
\eqno (B.11c)
$$

$\bullet h^{\theta\theta}$
$$
0= {h^{rr}\over r^4}{[2-(D-1)R_+-(3D-7)R_-+2(2D-5)R_+R_-]\over(1-R_+)(1-R_-)}
+{(D-3)h^{tt}\over r^4}{(R_+-R_-)(1-R_+)\over (1-R_-)}
$$
$$
+K \Bigl ( -\mu^2 - {\Omega^2(1-R_-)\over(1-R_+)}
+ {2\over r^2}[(D-2)+(D-4)(R_++R_-)-(D^2-3D-1)R_+R_-]
\Bigr )
$$
$$
-4f'(r){(1-R_+)(1-R_-)\over r^3}
+ [(D+2)-5(R_++R_-)-(D-8)R_+R_-]  {K'\over r}
+ (1-R_+)(1-R_-) K''
\eqno (B.11d)
$$

$\bullet h^{ti}$
$$
0=- {(D-3)\Omega (R_+ - 2R_- +R_+R_-)  \over  r(1-R_+)^2}  h^{ri}
-  {i\mu_i (D-3)R_-\over r(1-R_-)}  h^{tr}
+4i\mu_i \Omega f (r){(1-R_-)\over(1-R_+)}
$$
$$ 
-  h^{ti}\Bigl ( \mu^2 +{\Omega^2(1-R_-)\over(1-R_+)}\Bigr )
+{{h^{ti}}'\over r}(1-R_-)[(D-2)+(D-4)R_+]
+(1-R_+)(1-R_-){h^{ti}}''
\eqno (B.11e)
$$

$\bullet h^{ri}$
$$
0=- {(D-3)\Omega (R_+ - R_-)  \over  r}  h^{ti}
-  {i\mu_i (D-3)R_-\over r(1-R_-)}  h^{rr}
-4i\mu_i f'(r)(1-R_+)(1-R_-)
$$
$$ 
+  h^{ri}\Bigl ( -\mu^2 - {\Omega^2(1-R_-)\over(1-R_+)} +
{(1-R_+)[-(D-2)(1-(D-1)R_-)-(2D-5)R_-^2]\over r^2(1-R_-)}\Bigr )
$$
$$  
+{{h^{ri}}'\over r}(1-R_+)(D-2-(2D-5)R_-)
+(1-R_+)(1-R_-){h^{ri}}''
\eqno (B.11f)
$$

$\bullet h^{ij}$
$$
0=- {(D-3)R_- \over  r(1-R_-)} [i\mu_i h^{rj}+i\mu_jh^{ri}]
+4\mu_i \mu_j f (r)
-  h^{ij}\Bigl ( \mu^2 +{\Omega^2(1-R_-)\over(1-R_+)}\Bigr )
$$
$$ 
+{{h^{ij}}'\over r}[(D-2)-R_--R_+-(D-4)R_+R_-]
+(1-R_+)(1-R_-){h^{ij}}''
\eqno (B.11g)
$$

For the trace, h,
$$
0= -2(D-2)(D-3)^2R_+R_-K 
- \left( {\mu^2} + {\Omega^2( 1 - R_-) \over (1 - R_+)}\right)  
     (h-4f)
$$
$$
+{(1 - R_- )(D-2 - R_+)\over r}(h' -4f')
+{(1 - R_- )(1- R_+ )}(h'' -4f'')
\eqno (B.11h)
$$
and finally for the dilaton perturbation:
$$
0=- {(D-3)^2R_-(1-R_+)(R_+-R_-)\over4r^2(1-R_+)^2}h^{tt}
+{(D-3)(D-2)\over 2} KR_-[1-(D-2)R_+]
$$
$$
-{(D-3)R_-[2(D-2)-(3D-7)(R_++R_-)+2(2D-5)R_+R_-]\over 4r^2(1-R_-)^2(1-R_+)}h^{rr}
-  f\Bigl ( \mu^2 +{\Omega^2(1-R_-)\over(1-R_+)}\Bigr )
$$
$$ 
+{f'\over r}[(D-2)-R_++(D-4)R_--(2D-7)R_+R_-]
+(1-R_+)(1-R_-)f''
\eqno (B.11i)
$$
\vfill\eject

\proclaim Captions.

\noindent{\bf Figure 1.} Penrose diagram for the metric of eq.(1).  All but 
two dimensions have been suppressed; each point of this diagram corresponds to  
a D-2 sphere times 10-D flat space.  The line at $v=v_0$ is the initial 
hypersurface on which the perturbations are set.  It is a Cauchy surface 
for the spacetime exterior of the black hole.
\smallskip

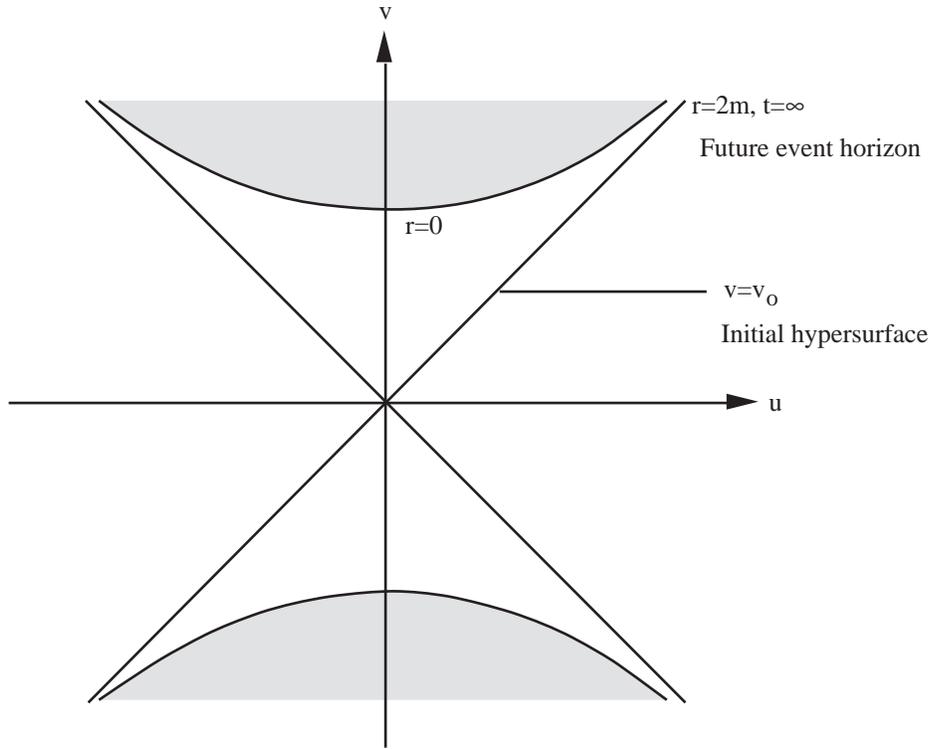
\noindent{\bf Figure 2.} Plot of the modes $H^{tt},H^{tr}$ and $H^{rr}$
as a function of $(r-r_+)$ for a regular mode.  As $(r-r_+)$ becomes
large all the functions decay exponentially, they reach a maximum at small $(r-r_+)$
and go to zero as $(r-r_+)\rightarrow 0$.
\smallskip

\noindent{\bf Figure 3.} Plot of $\Omega$ as a function of $\mu$ for black 
strings and branes with $D=4,...,9$ and $r_+=2$ for which an instability has 
been found. 
The bold points correspond to value calculated numerically and the
lines have been traced to guide the eye. 
\smallskip

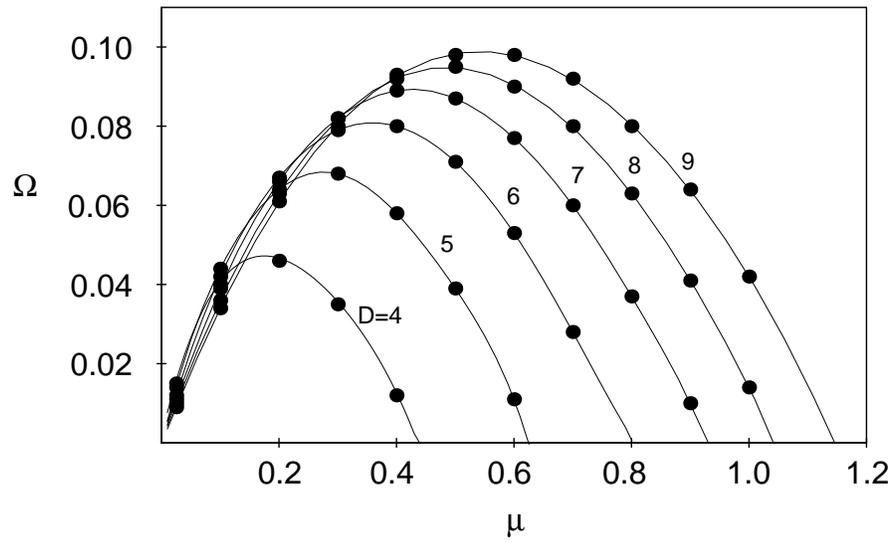
\noindent{\bf Figure 4.}  Schematic diagram of the instability.  All the dimensions 
have been suppressed except the $r$ and $z$ one.  The geometry
is initially invariant under a $z$ translation.  The instability increases
the size of the apparent horizon at some values of $z$ and decreases it
at other values.  In the chargeless case it seems reasonable that the 
apparent horizon will brake into different parts.
\smallskip

\noindent{\bf Figure 5.} Plot of the modes $f,q, H^{zz}$ and $H^{tz}$
as a function of $(r-r_+)$ for a regular mode.  As $(r-r_+)$ becomes
large all the functions decay exponentially, they reach a maximum at small $(r-r_+)$ 
and go to zero as $(r-r_+)\rightarrow 0$.
\smallskip

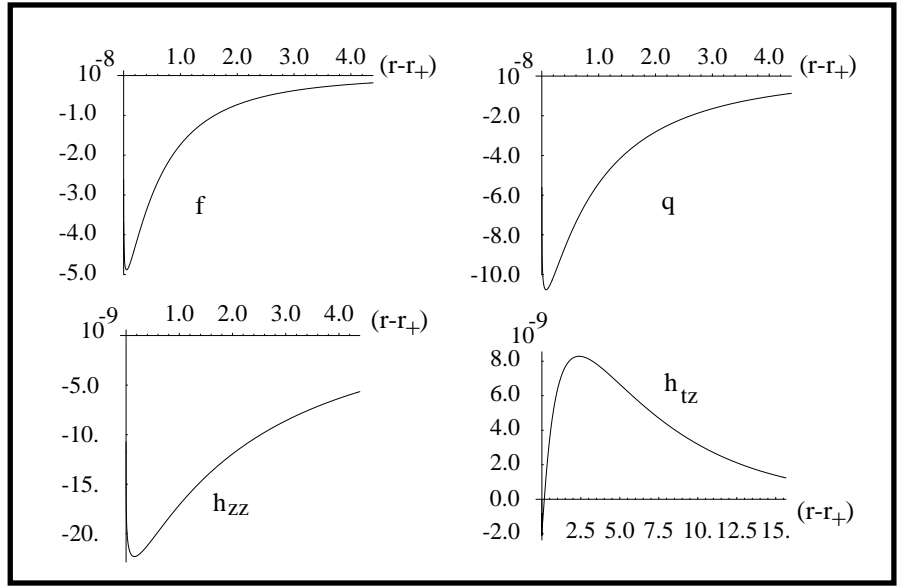
\noindent{\bf Figure 6.} Plot of $\Omega$ as a function of $\mu$ for a charged 
5d black string with charges corresponding to $r_-=0,1.0,1.5$ and $r_+=2$ for 
which an instability has been found. 
The bold points correspond to value calculated numerically and the
lines have been traced to guide the eye. 
\smallskip

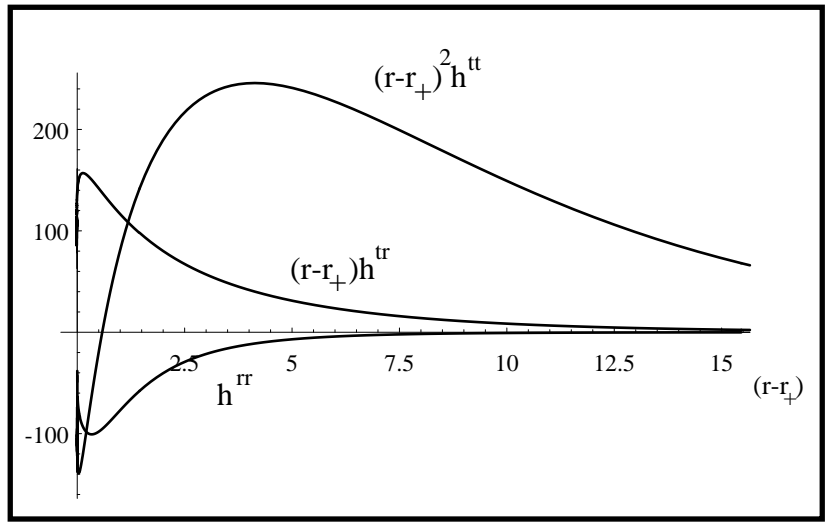
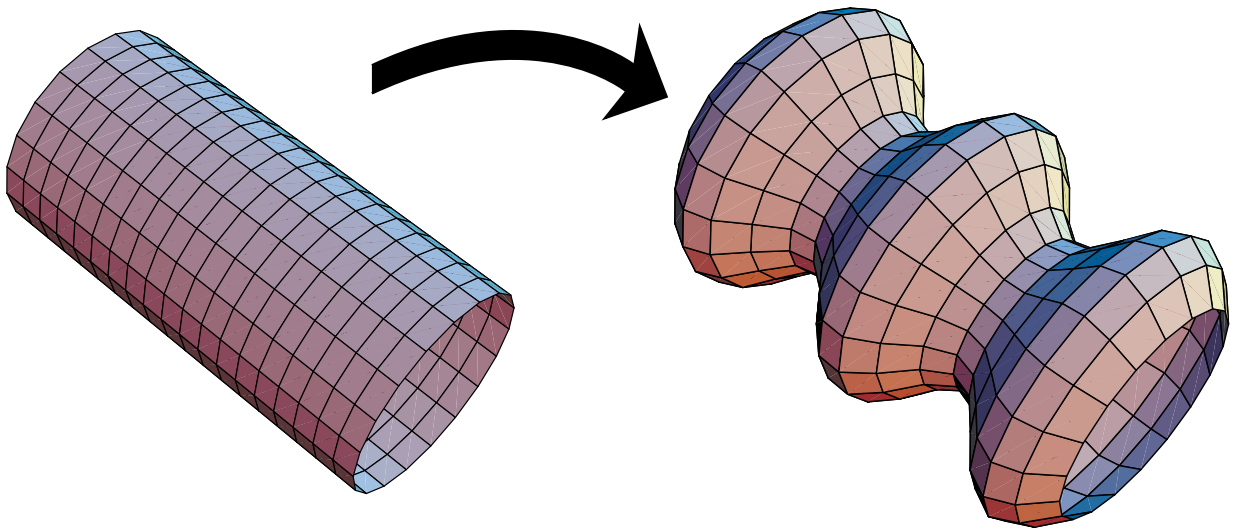
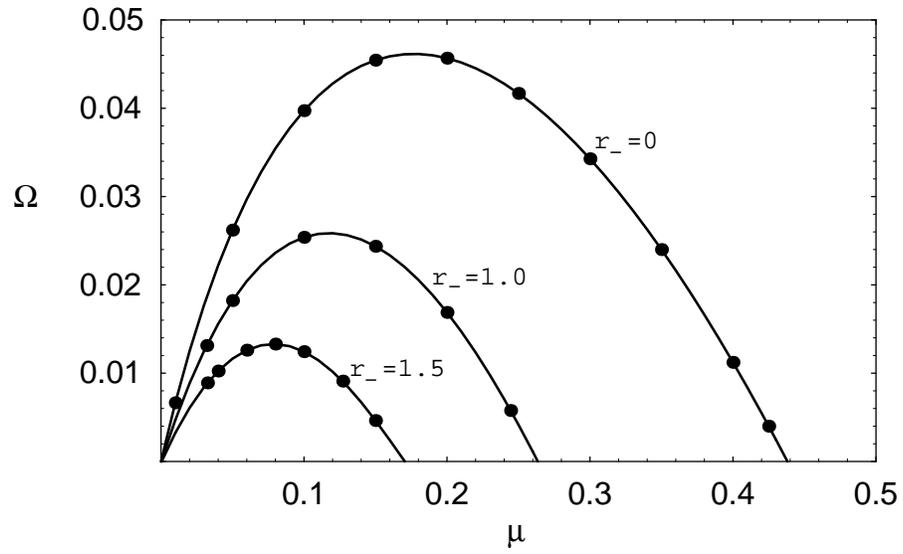
\end